\documentclass[]{aastex63}

\newcommand\bmm{{$\beta$-meteoroid }}
\newcommand\bms{{$\beta$-meteoroids }}
\newcommand\amsn{{$\alpha$-meteoroids}}
\newcommand\amm{{$\alpha$-meteoroid }}
\newcommand\ams{{$\alpha$-meteoroids }}
\newcommand\bmsn{{$\beta$-meteoroids}}
\newcommand\bs{{$\beta$-stream }}
\newcommand\bsn{{$\beta$-stream}}
\newcommand\bss{{$\beta$-streams }}
\newcommand\bssn{{$\beta$-streams}}
\newcommand\rs{R_\odot}
\newcommand\rst{$\rs$}
\newcommand\isois{IS$\odot$IS }

\newcommand\kms{km s$^{-1}$}
\newcommand\kgs{kg s$^{-1}$}

\usepackage{mathtools}
\usepackage{amsmath}
\usepackage{amssymb}
\usepackage{rotating}
\usepackage{latexsym}

\received{21-Jan-2021}
\revised{15-Apr-2021}
\accepted{16-Apr-2021}
\submitjournal{PSJ}

\shorttitle{PSP Dust: Orbits 1-6}
\shortauthors{Szalay et al.}
\graphicspath{{./}{figures/}}

\begin{document}

\title{Collisional Evolution of the Inner Zodiacal Cloud}

\correspondingauthor{Jamey Szalay}
\email{jszalay@princeton.edu}

\author[0000-0003-2685-9801]{J. R. Szalay}
\affil{Department of Astrophysical Sciences, Princeton University, 4 Ivy Ln., Princeton, NJ 08540, USA}

\author[0000-0002-5667-9337]{P. Pokorn\'y}
\affiliation{Astrophysics Science Divison, NASA Goddard Spaceflight Center, Greenbelt, MD, 20771, USA}
\affiliation{Department of Physics, The Catholic University of America, Washington, DC, 20064, USA}

\author[0000-0003-1191-1558]{D. M. Malaspina}
\affiliation{Department of Astrophysical and Planetary Sciences, University of Colorado Boulder, Boulder, CO, USA}
\affiliation{Laboratory for Atmospheric and Space Physics, University of Colorado Boulder, Boulder, CO, USA}

\author[0000-0002-3081-8597]{A. Pusack}
\affiliation{Laboratory for Atmospheric and Space Physics, University of Colorado Boulder, Boulder, CO, USA}

\author[0000-0002-1989-3596]{S. D. Bale}
\affiliation{Space Sciences Laboratory, University of California, Berkeley, CA, USA}
\affiliation{Physics Department, University of California, Berkeley, CA, USA}

\author[0000-0002-8692-6925]{K. Battams}
\affiliation{US Naval Research Laboratory, 4555 Overlook Avenue, SW, Washington, DC, USA}

\author[0000-0002-9000-7630]{L. C. Gasque}
\affiliation{Space Sciences Laboratory, University of California, Berkeley, CA, USA}
\affiliation{Physics Department, University of California, Berkeley, CA, USA}

\author[0000-0003-0420-3633]{K. Goetz}
\affiliation{School of Physics and Astronomy, University of Minnesota, Minneapolis, MN, USA }

\author[0000-0003-1488-0283]{H. Kr\"uger}
\affiliation{Max-Planck-Institut f\"ur Sonnensystemforschung, G\"ottingen, Germany}

\author[0000-0001-6160-1158]{D. J. McComas}
\affil{Department of Astrophysical Sciences, Princeton University, 4 Ivy Ln., Princeton, NJ 08540, USA}

\author[0000-0002-3737-9283]{N. A. Schwadron}
\affiliation{University of New Hampshire, Durham, NH, USA}

\author{P. Strub}
\affiliation{Max-Planck-Institut f\"ur Sonnensystemforschung, G\"ottingen, Germany}




\begin{abstract}

The zodiacal cloud is one of the largest structures in the solar system and strongly governed by meteoroid collisions near the Sun. Collisional erosion occurs throughout the zodiacal cloud, yet it is historically difficult to directly measure and has never been observed for discrete meteoroid streams. After six orbits with Parker Solar Probe (PSP), its dust impact rates are consistent with at least three distinct populations:  bound zodiacal dust grains on elliptic orbits (\amsn), unbound \bms on hyperbolic orbits, and a third population of impactors that may either be direct observations of discrete meteoroid streams, or their collisional byproducts (``\bsn s''). \bss of varying intensities are expected to be produced by all meteoroid streams, particularly in the inner solar system, and are a universal phenomenon in all exozodiacal disks. We find the majority of collisional erosion of the zodiacal cloud occurs in the range of $10-20$ solar radii and expect this region to also produce the majority of pick-up ions due to dust in the inner solar system. A zodiacal erosion rate of at least {\color{black}$\sim$100 kg s$^{-1}$} and flux of \bms at 1 au of {\color{black}$0.4-0.8 \times 10^{-4}$ m$^{-2}$ s$^{-1}$} is found to be consistent with the observed impact rates. The \bms investigated here are not found to be primarily responsible for the inner source of pick-up ions, suggesting nanograins susceptible to electromagnetic forces with radii below $\sim$50 nm are the inner source of pick-up ions. We expect the peak deposited energy flux to PSP due to dust to increase in subsequent orbits, up to 7 times that experienced during its sixth orbit.


\end{abstract}



\section{Introduction} \label{sec:intro}

The zodiacal dust distribution in the inner solar system is continuously evolving. Grains orbiting the Sun slowly lose their angular momentum due to Poynting-Robertson drag \citep{burns:79a} and spiral in toward the Sun. Additionally, cometary and asteroidal activity inputs dust into the zodiacal cloud. Collisions occur between grains within the zodiacal cloud, which fragment into numerous smaller grains \citep{kuchner:10a}. If these collisional products are sufficiently small, they experience enough repulsive force from radiation pressure to overcome the Sun's attractive gravitational force and leave the solar system on unbound, hyperbolic trajectories. Such grains are called \bmsn, after the parameter $\beta$, the ratio of radiation pressure to solar gravity \citep{zook:75a}. 

Dust populations in the inner solar system have been observed both directly via dedicated dust instruments, and indirectly, typically via signals produced on spacecraft antenna. Dedicated dust measurements in the inner solar system have been made with Pioneers 8 and 9 \citep{berg:73a,grun:73a}, HEOS-2 \citep{hoffmann:75a, hoffmann:75b}, Helios \citep{leinert:78a, leinert:81a, grun:80a, altobelli:06a, kruger:20a}, Ulysses \citep{wehry:99a,wehry:04a,landgraf:03a, sterken:15a, strub:19a}, and Galileo \citep{grun:97a}. These observations identified three populations of dust:  larger grains on bound orbits (\amsn), smaller grains on unbound orbits (\bmsn), and interstellar grains transiting the solar system.

The Parker Solar Probe (PSP) mission \citep{fox:16a} transits the inner-most portions of our solar system's zodiacal dust cloud. During its first orbit, PSP flew by the Sun with a perihelion distance of 0.17 au (36 solar radii). Throughout the mission, PSP's perihelion distance is incrementally decreased via Venus flybys \citep{guo:21a}. The Solar Probe Science Definition Team identified the exploration of ``dusty plasma phenomena and their influence on the solar wind and energetic particle formation'' as one of its four main science objectives \citep{mccomas:07a}. Many of the questions to be resolved for this science objective rely on making measurements with a dedicated dust analyzer with composition capability. While unequipped with a dedicated dust detector, PSP registers dust impacts primarily via potential measurements with the FIELDS instrument \citep{bale:16a}. In lieu of measurements with a dedicated dust detector, these observations provide the best in-situ characterization of the dust environment inside 0.3 au to date.  There has been an extensive history of measuring dust impacts with spacecraft antenna measurements, for example: Voyager 2 \citep{gurnett:83a}, Vega \citep{laakso:89a}, DS-1 \citep{tsurutani:03a,tsurutani:04a}, Wind \citep{malaspina:14a,kellogg:16a,malaspina:16a}, MAVEN \citep{andersson:15a}, STEREO \citep{zaslavsky:12a,malaspina:15a} and MMS \citep{vaverka:18a,vaverka:19a}.

During its first three orbits, impact rates observed by PSP were consistent with fluxes of high-speed, submicron-sized \bms leaving the solar system on escaping orbits \citep{szalay:20a,page:20a,malaspina:20a}. Nano-sized grains which experience strong electromagnetic forces were not found to be dominantly responsible for the impact rates during PSP's 2nd orbit \citep{mann:20a}. After six orbits spanning three distinct orbital families (1-3, 4-5, 6), the PSP impact rate data revealed a more complex picture than garnered from the first three orbits. In this study, we present the first six orbits of PSP FIELDS impact rate data in Section~\ref{sec:rates}. We introduce a two-component dust model and compare its results to the data in Section~\ref{sec:model}, showing how it reproduces the broad features in the data. In Section~\ref{sec:enhance}, we present two possibilities for an anomalous feature in the data not described by the two-component model. We discuss implications for the upcoming orbits in Section~\ref{sec:predict} and conclude with a discussion of these results in Section~\ref{sec:discuss}.

\section{Impact Rates for Orbits 1-6} \label{sec:rates}

\subsection{Dust identification and noise determination}

{\color{black}The PSP Fields Experiment (FIELDS) \citep{bale:16a} detects dust impacts on the spacecraft surface.  Dust grains traveling at high relative velocities ($>$ 1 km/s) to the spacecraft vaporize and ionize upon impact, creating a transient plasma cloud detected by the FIELDS electric field antennas as a high amplitude (10's to 1000's of mV), impulsive (10's of $\mu$s) voltage spike.  Since dust is a fundamental part of the ambient near-Sun environment, and because FIELDS is sensitive to the voltage perturbations induced by dust impacts, FIELDS unavoidably measures dust. Since the entire spacecraft surface area acts as part of the `detector', antenna-based dust detection has the advantage of high count rates.   PSP is bombarded by thousands of such impacts each orbit, creating a statistically robust data set of dust detections \citep[e.g.][]{page:20a,malaspina:20a,szalay:20a}.

FIELDS has four 2 m whip antennas in the plane of PSP's heat shield (V1, V2, V3, V4), and a fifth, 21 cm antenna situated along the magnetometer boom (V5). FIELDS dust counts are most readily determined in peak detection data from the Time Domain Sampler (TDS) receiver. The TDS records signals from each antenna in a bandpass from a few kHz to 1 MHz. It produces a data product called TDSmax, which reports the largest (signed) amplitude value on each configured channel recorded by the TDS over a short time interval.  The TDSmax data product also records the number of zero-crossings associated within $\sim$13 ms surrounding the time of the largest-amplitude signal.  This information is used to distinguish dust impact voltage spikes (few zero crossings) from plasma waves (many zero crossings).  

Voltage spikes produced by impulsive dust impact voltage signals are clearly distinguished from plasma wave-induced signals in TDSmax data by amplitude $>$50 mV and $<$5 zero crossings \citep{page:20a}. Since PSP's first solar encounter in 2018, TDSmax data for the V2 monopole channel have been continuously produced whenever FIELDS is powered on, producing a vast database of dust impacts. In this work, we analyze impacts registered by FIELDS above 50 mV identified in the V2 TDSmax dataset. In addition to count rates, FIELDS can determine basic directionality information for each dust impact by comparing voltage spike signal strength across its multiple antennas \citep{malaspina:20a}. The time domain waveform shapes of these voltage spikes, determined via high-resolution waveform captures, are consistent with previous observations of dust impact voltage spikes \citep{mann:19a,bale:20a}. }


Changes in solar wind plasma density and temperature cause the spacecraft floating potential to vary and could affect dust impact observations.  Because the amplitude of measured dust voltage spikes is partially determined by the amount of impact-liberated charge recollected on spacecraft surfaces, variations in the spacecraft floating potential will produce some variation in the minimum detectable dust impact voltage spike amplitude, and therefore variation in the dust count rate.  However, the spacecraft floating potential has varied weakly over orbits 1-6, with most excursions less than 5 Volts in magnitude \citep{bale:20a}.  Therefore, we do not expect the impact count rate to vary significantly over any given orbit due to spacecraft charging effects when a fixed amplitude threshold is used. 

The identified dust impact rates have been consistent with a ``memory-less'' processes, such that the probability of receiving an impact is independent of the probability of when the next impact will occur (Poissonian). While there may be additional non-Poissonian sources of similar waveforms \citep{mozer:20a}, the finding that PSP's identified dust impact rates are highly consistent with a Poisson process, along with their waveform similarity to known dust impact signals, reinforces that the vast majority of observed abrupt voltage spikes are due to dust impacts \citep{page:20a}.

\subsection{Impact rates}

\begin{figure}
\plotone{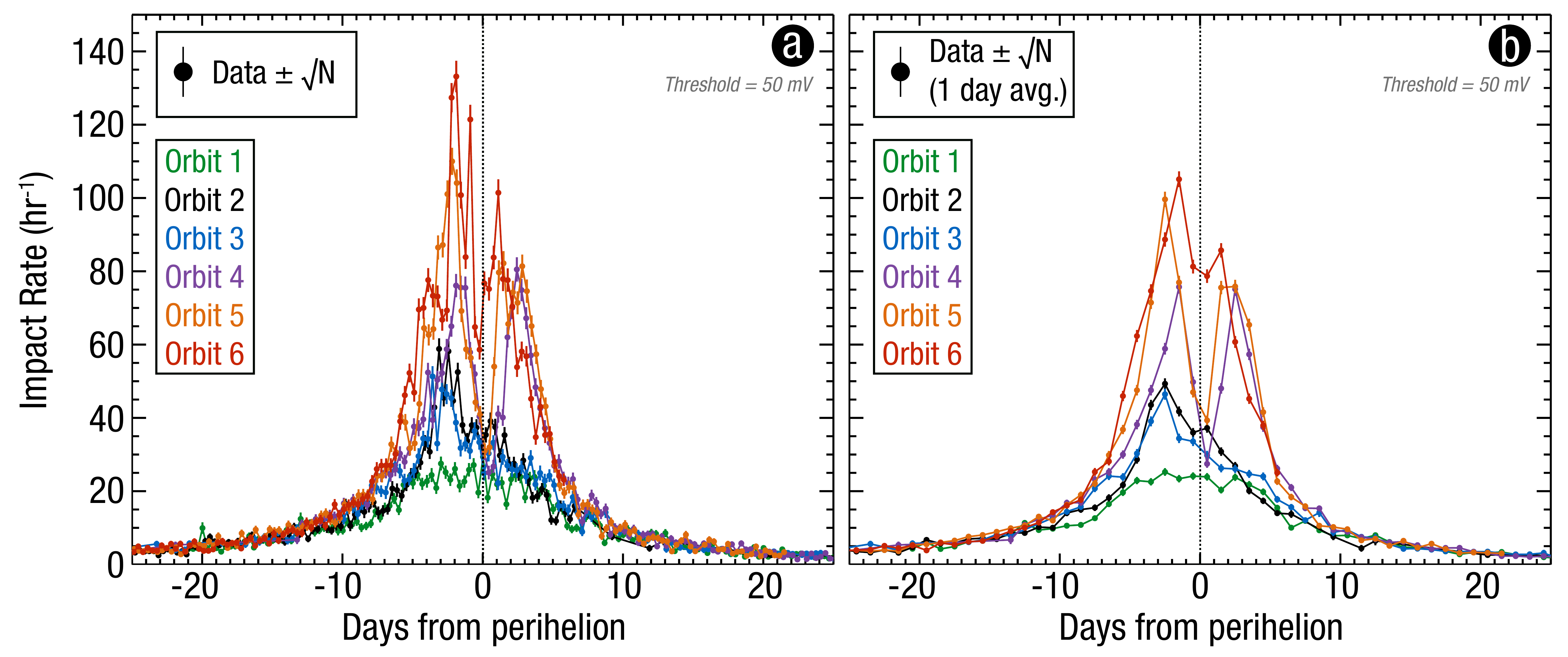}
\caption{Impact rates as a function of days from perihelion for orbits 1-6 (color coded in the legend) with an impact threshold of 50 mV for high temporal resolution (a) and 1 day bins (b). Error bars show $\sqrt{N}$ counting statistics.  \label{fig:rates}}
\end{figure}

\begin{figure}
\plotone{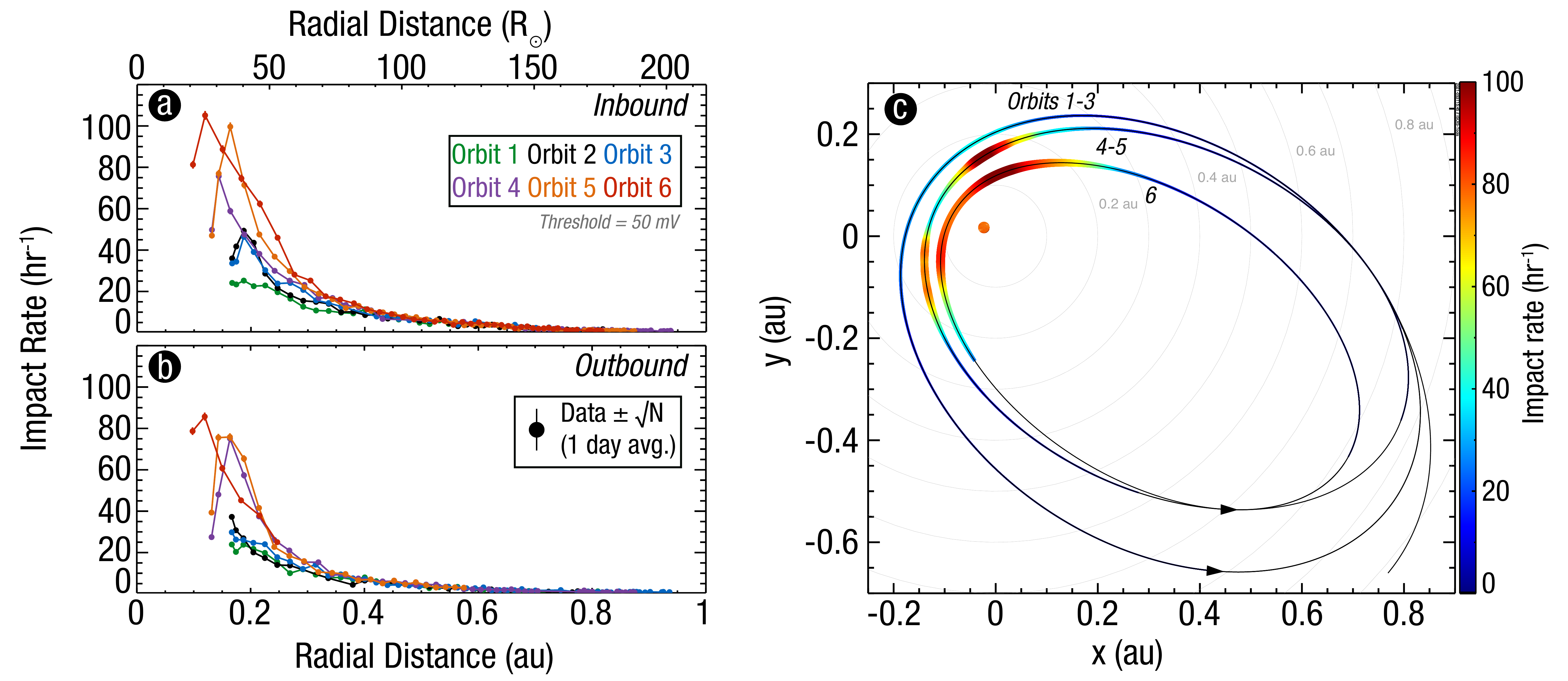}
\caption{Daily averaged impact rates as a function of radial distance for orbits 1-6, separated by inbound (a) and outbound (b). Error bars show $\sqrt{N}$ counting statistics, {\color{black}where due to the large number of counts the error bars are often smaller than the plotting symbols themselves}. (c) Impact rates overlaid on the PSP trajectory in the ecliptic J2000 frame, averaged over orbits 1-3, 4-5, and individually shown for orbit 6. Color and width of the color strip represents the impact rate. \label{fig:rates_overview}}
\end{figure}

Figure~\ref{fig:rates}a  shows the impact rates observed by the FIELDS instrument on PSP for each orbit in high resolution for a voltage threshold of 50 mV. Errors are given for $\sqrt{N}$ counting statistics, where $N$ is the total number of impacts in each measurement. To discern the dust populations responsible for these impact rates, we seek to understand the large-scale, broad features in the data. Hence, we apply a 1 day time binning to the data, shown in Figure~\ref{fig:rates}b; this work focuses on these time-binned data. Figure~\ref{fig:rates_overview}a,b shows the rates as a function of the distance from the center of the Sun. Figure~\ref{fig:rates_overview}c shows color-coded trajectories according to the impact rates, where orbits 1-3, and 4-5 have been averaged together, and orbit 6 is shown individually. 

There are multiple features of note across each of these six orbit in Figure~\ref{fig:rates_overview}. All orbits exhibit a peak on the inbound, pre-perihelion portion of the orbital arc. Outbound, post-perihelion, orbits 4 and 5 show a clear second peak in their impact rate profiles, while the remaining have an extended ``shoulder'' that generally decreases monotonically from the pre-perihelion peak. The location of the first peak occurs at similar longitudes/true anomalies across the six orbits. We seek to explain these various features and compare the observed impact rates to a dust model described in the next section.

\section{Two-component Dust Model} \label{sec:model}

\subsection{Dust populations and terminology}
There have been two distinct dust populations observed by in-situ detection methods that are directly related to our zodiacal cloud. The bulk of the material in the zodiacal cloud is comprised of meteoroids on bound, elliptic orbits slowly spiraling into the Sun under the influence of Poynting-Robertson drag \citep[e.g.][]{grun:85a}. Grains with these orbital characteristics have been termed \ams \citep{grun:80b}. These grains collide and fragment into smaller grains, which can experience sufficient force from radiation pressure to be gravitationally unbound from the Sun. Once unbound, grains move along hyperbolic orbits, with trajectories that are governed by $\beta$, the ratio of solar radiation and gravitational forces \citep[e.g.][]{zook:75a,burns:79a}. Hence, these grains are termed \bmsn. We incorporate both \ams and \bms into a two-component model to compare with PSP impact rates. While there may also be a minor contribution of intersteller grains \citep[e.g.][]{grun:93a,sterken:12a,strub:19a}, we do not model this population as their fluxes are expected to be particularly small given their exclusion from the inner solar system due to radiation pressure.

\subsection{Impact Charge Generation}

Dust impacts into a target (PSP) generate an impact charge according to the relation,
\begin{equation}
  Q_\mathrm{imp} = Cm_\mathrm{imp}^av_\mathrm{imp}^b,
  \label{eq:qimp}
\end{equation}
where $m_\mathrm{imp}$ is the impactor mass, $v_\mathrm{imp}$ is the impactor speed, and $C$, $a$, and $b$ are empirically determined constants based on the impactor and target properties \citep{auer:01a}. {\color{black}These parameters can vary for different surfaces, and the constant $C$ has been experimentally determined to span a large range \citep{collette:14b}. We will perform our fits, described in the next sections, on $\tilde{Q}_\mathrm{imp} \equiv Q_\mathrm{imp}/C$ to remove any dependence on this coefficient.} Without a specific calibration for the various surfaces and target materials on PSP, we assume typical values of $a = 1$, and $b = 3.5$ \citep{auer:01a} for the calculations  and analysis below, where $m_\mathrm{imp}$ is in kg and $v_\mathrm{imp}$ in km s$^{-1}$. Any detection method based on the amount of impact charge generated, in this case via the capacitively coupled transient voltage changes measured by FIELDS, will have a critical value of impact charge below which an impact creates too little charge to detect, $Q_c$. For any given impact speed, $Q_c$ sets the minimum detectable radius via
\begin{equation}
  s_{min} = \left(  \frac{3Q_c}{2\pi\rho v_\mathrm{imp}^{b}} \right)^{1/3} = C_a v_\mathrm{imp}^{-b/3},
  \label{eq:a_min}
\end{equation}
where $\rho$ is the bulk mass density of the impactors and $C_a = (3Q_c / 2\pi\rho)^{1/3}$. We have left $b=3.5$ as a parameter to show the dependence on this exponent in subsequent equations.  To roughly estimate the expected impact charge threshold, we assume PSP is spherical with a capacitance of $C_\mathrm{PSP} = 200$ pF. The voltage amplitude generated by a dust impact is given by $Q_\mathrm{imp} =  C_\mathrm{PSP}\Delta V/\Gamma_q$ where $\Delta V$ is the potential change and $\Gamma_q$ is the efficiency factor, typically $\sim$0.2 \citep{collette:15a,collette:16a}. With a detection threshold of {\color{black}$\Delta V_c = 50$ mV} and $\Gamma_q = 0.2$, the threshold impact charge is estimated to be {\color{black}$Q_c \approx 0.3 \times 10^9$} electrons.


\subsection{Bound $\alpha$-meteoroids on elliptic orbits}

We use the following functional form for the number density $n_\alpha$ of bound \ams on elliptic orbits,
\begin{equation}
n_\alpha(r,a) = n_{0\alpha} f_\alpha(r) r^{-1.3} \left( \frac{s}{s_0} \right)^{-3\alpha}
\label{eq:na}
\end{equation}
where $n_{0\alpha}$ is the cumulative number density at 1 au for grains with radii greater than the reference grain size $s_0 = 1$ $\mu$m, approximating the size distribution using a single power-law, and $r$ is the heliocentric distance in au. Mass indices varying from 0.4 to 1.3 were observed over a large mass range at 1 au \citep{grun:85a} and modeling efforts suggest the size distribution is expected to differ from this inside 1 au \citep{ishimoto:98a}. To reduce model parameters, we assume a single power-law dependence with $\alpha=0.9$, consistent with the size index for $\sim$100 $\mu$m grains at 1 au \citep{grun:85a,pokorny:16a} and with the collisionally produced size distributions assumed for the meteoroid stream analysis discussed in this work (Appendix~\ref{app:rate}). $f_\alpha(r)$ is given by the following empirical relation \citep{stenborg:20a} determined via fits to the Wide-field Imager for Parker Solar Probe (WISPR) remote sensing data \citep{vourlidas:16a},
\begin{equation}
f_\alpha(r) = \left\{
    \begin{array}{rl}
      0 & \ \  \text{if } r < r_0, \\
      \frac{r-r_0}{r_1-r_0} & \ \ \text{if } r_0 \leq r \leq r_1, \\
      1 & \ \ \text{if }  r > r_1.
     \end{array} 
   \right.
   \label{eq:zody_den}
\end{equation}
Equation~\ref{eq:zody_den} matches the Helios remote observations outside $r>r_1$ \citep{leinert:78b} and provides an estimated representation for how the density would trend inside $r_1$ towards the dust free zone \citep[e.g.][]{leinert:78a,stenborg:20a} approximated here to occur at $r_0 = 5\rs$.  We use a value of $r_1 = 19\rs$ as empirically determined by comparison with WISPR data \citep{stenborg:20a}, which is inside the perihelion distance for orbits 1-6 discussed in this work. All grains are assumed to be on perfectly circular orbits with orbital speeds of $v_{orb} = \sqrt{\mu/r}$, where $\mu = 1.327 \times 10^{20}$ m$^3$ s$^{-2}$ is the Sun's gravitational parameter. At any given location along PSP's orbit, the observed zodiacal impact flux is,
\begin{equation}
\Phi_\alpha(r) = n_{0\alpha}f_\alpha(r)(s_0/C_a)^{3\alpha}v_\mathrm{imp}^{1+\alpha b}r^{-1.3} .
\end{equation} 
The impact speed between \ams and PSP is given by \citep[e.g.][]{szalay:20a},
\begin{equation}
v_\mathrm{imp}^2 = \mu \left(  \frac{3}{r} - \frac{1}{a} - 2\sqrt{ \frac{a(1-e^2)}{r^3} } \right),
\end{equation}
where $a$ and $e$ are Parker Solar Probe's semi-major axis and eccentricity, respectively, and PSP and all grains are assumed to have inclination $i=0$. For bound grains on circular orbits, we assume they have the density and $\beta$ characteristics of nominal old cometary grains \citep{wilck:96a}, such that grains with radii above 0.5 $\mu$m have $\beta < 0.5$ and are bound.

\subsection{Unbound \bms on hyperbolic orbits}

Initial analyses assumed all \bms were produced at a single location and had the same $\beta$ value \citep{szalay:20a,page:20a,malaspina:20a}. Here, we utilize a forward-model to track \bmm trajectories from a dispersed source region with a distribution of masses and $\beta$ values. We model the dynamics of \bms under the attractive solar gravitational force and repulsive solar radiation force: $F = \mu(1-\beta)/r^2$. Grains collisionally produced at perihelion of their parent body's orbit will be unbound from the Sun's gravity for $\beta \geq 0.5(1-e)$ where $e$ is the eccentricity of the parent body \citep{grun:85a}. For example, \bms produced from a parent grain on a perfectly circular orbit are unbound for $\beta \geq 0.5$. The value of $\beta$ is dependent on the grain composition, albedo, porosity, and size. We assume collisional products are similar to canonical ``young'' cometary grains \citep{wilck:96a} to determine the relation between mass and $\beta$. Assuming collisions in the nominal zodiacal cloud occur between grains on perfectly circular orbits, collisional products with radii less than 1.2 $\mu$m are unbound and become \bmsn.  

We estimate PSP impact rates using the following numerical scheme. In a uniform grid of 100 bins in radial position from 0 to 1 au, we release \bms and track their location and speed as they travel away from the Sun. The simulation is performed in 1D, assuming PSP and all grain trajectories are in the same plane. From their radial speed, we accumulate residence times for grains released at all initial positions. We assume the mass distribution follows a power-law with differential slope $\alpha=0.9$ (Appendix~\ref{app:rate}) and follow grain trajectories within 50 logarithmically spaced mass bins corresponding to a radius range of 50 nm to 1.2 $\mu$m. We normalize residence times from grain trajectories according to the size and collisional mass production distribution to account for which initial starting locations and masses contribute to the density at any location. 

To approximate the functional form of collisional mass production, we follow previous conventions \citep{zook:75a,steel:86a}, by assuming the collision rate per unit distance is proportional to $n_\alpha(r)^2v(r)\sigma(r)$, where $n_\alpha$ is the spatial zodiacal density given in Equation~\ref{eq:na}, $v$ is the average impact speed, and $\sigma$ is the cross sectional area. The average impact speed is assumed to follow the Keplerian trend proportional to $r^{-0.5}$. Taking into account the threshold energy to catastrophically destroy grains (Appendix~\ref{app:rate}, Eq.~\ref{eq:sc}), we approximate the cross-sectional area to be proportional to $r^{-\alpha+2/3}$ . Therefore, we represent the differential collisional mass production as,
\begin{equation}
M^+(r)dV \propto n_\alpha(r)^2v(r)\sigma(r)4\pi r^2 dr \propto f^2_z r^{-1.33},
\end{equation}
where the integral over $\int M^+(r)dV = M_\mathrm{tot}$ is a fit parameter. The density of \bms detectable by PSP is then $n_\beta(r,a) \propto n_\beta(r) s_{min}^{-3\alpha} \propto n_\beta v^{\alpha b}$. The incident impact flux of \bms observed by PSP is  
\begin{equation}
\Phi_\beta = n_\beta(r,a)v(r) \propto n_\beta v_{imp}^{1+\alpha b} .
\end{equation}

\subsection{Model fits}

For any combination of the three fit parameters ($n_{0\alpha}$, $\tilde{Q}_c$, and $M^+_\mathrm{tot}$), the model can produce synthetic impact rates to the PSP spacecraft. The fits are performed using a $\chi^2$ minimization \citep{markwardt:09a} between synthetic forward-modeled impact rates and the rates observed by PSP. We individually fit each orbit, with fit values given in Table~\ref{table:fit}. 

\begin{table}
\begin{centering}
{\color{black}\begin{tabular}{cccc|c}
Orbit & $n_{0\alpha}$ & $\tilde{Q}_c$ & $M^+_\mathrm{tot}$ & $\Phi_\beta$(1 au) \\
 & (km$^{-3})$ & $(10^9$ e$^-$) & (kg s$^{-1}$) & $10^{-4}$ m$^{-2}$ s$^{-1}$ \\
\hline \hline
1 & 0.3 & 1.4 & 60 & 0.4 \\
2 & 0.1 & 1.0 & 110 & 0.6\\
3 & 0.3 & 1.8 & 100 & 0.6 \\
4 & 0.6 & 3.4 & 85 & 0.5 \\
5 & 0.6 & 3.4 & 110 & 0.6 \\
6 & 0.7 & 6.6 & 100 & 0.8 \\
\hline
1-6 & 0.5 & 3.4 & 100 & 0.6 \\ 
\hline \hline
\end{tabular}}

\caption{\label{table:fit}Two-component fit values for individual orbits and all orbits together. Columns 2-4 give the model fits directly from the two-component model, while the last column ($\Phi_\beta$) gives the expected flux of \bms at 1 au.}
\end{centering}
\end{table}

\begin{figure}[ht!]
\plotone{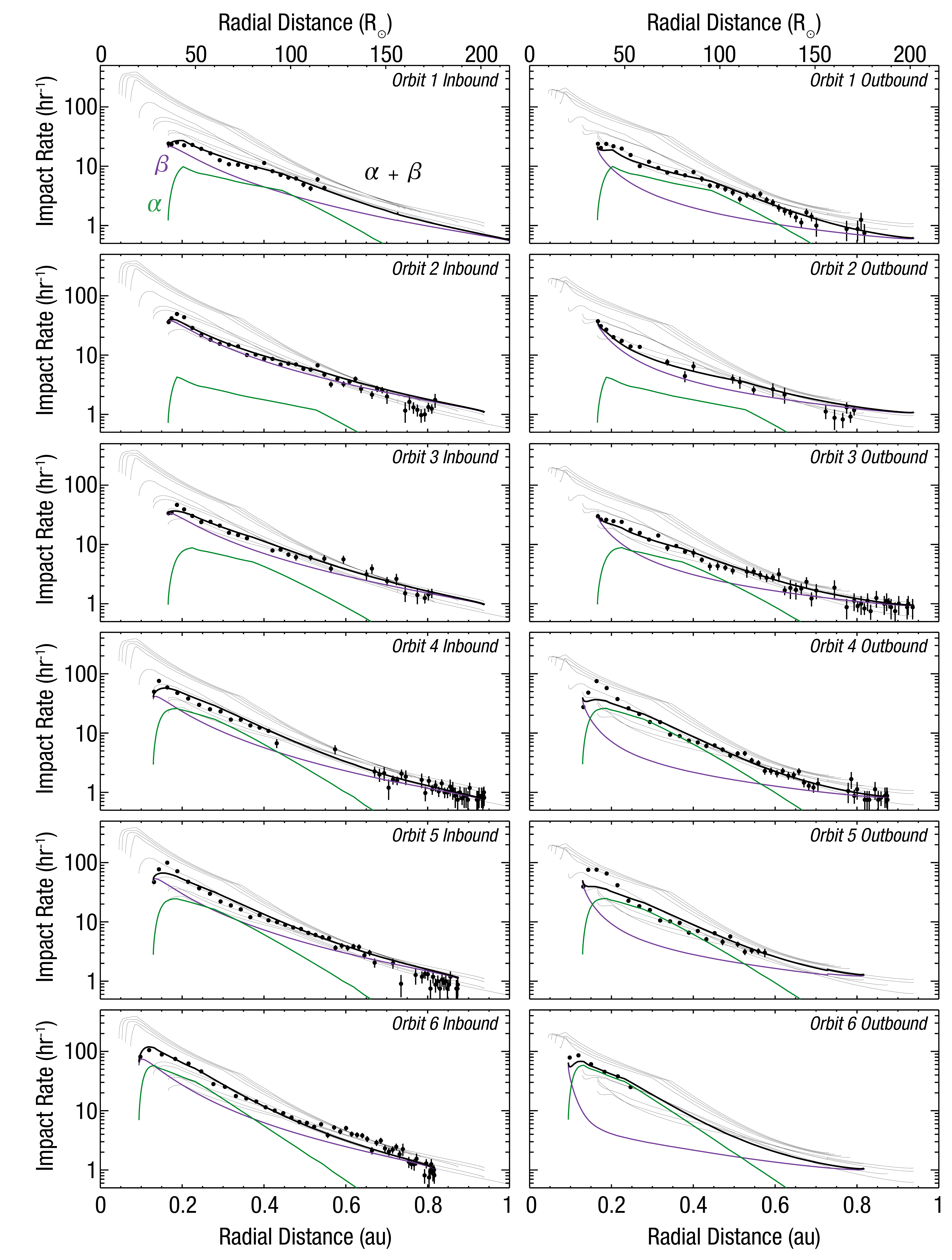}
\caption{Model fits for each orbit for inbound (pre-perihelion) and outbound (post-perihelion) arcs.  {\color{black}Individual PSP data shown as the black dots. The individual \amm and \bmm fits are shown in the green and purple lines respectively, where their sum is given by the black overall impact rate fit. Fits for all PSP orbits up to orbit 24 are duplicated in each panel, shown with grey lines.} \label{fig:radial}}
\end{figure}

\begin{figure}[ht!]
\plotone{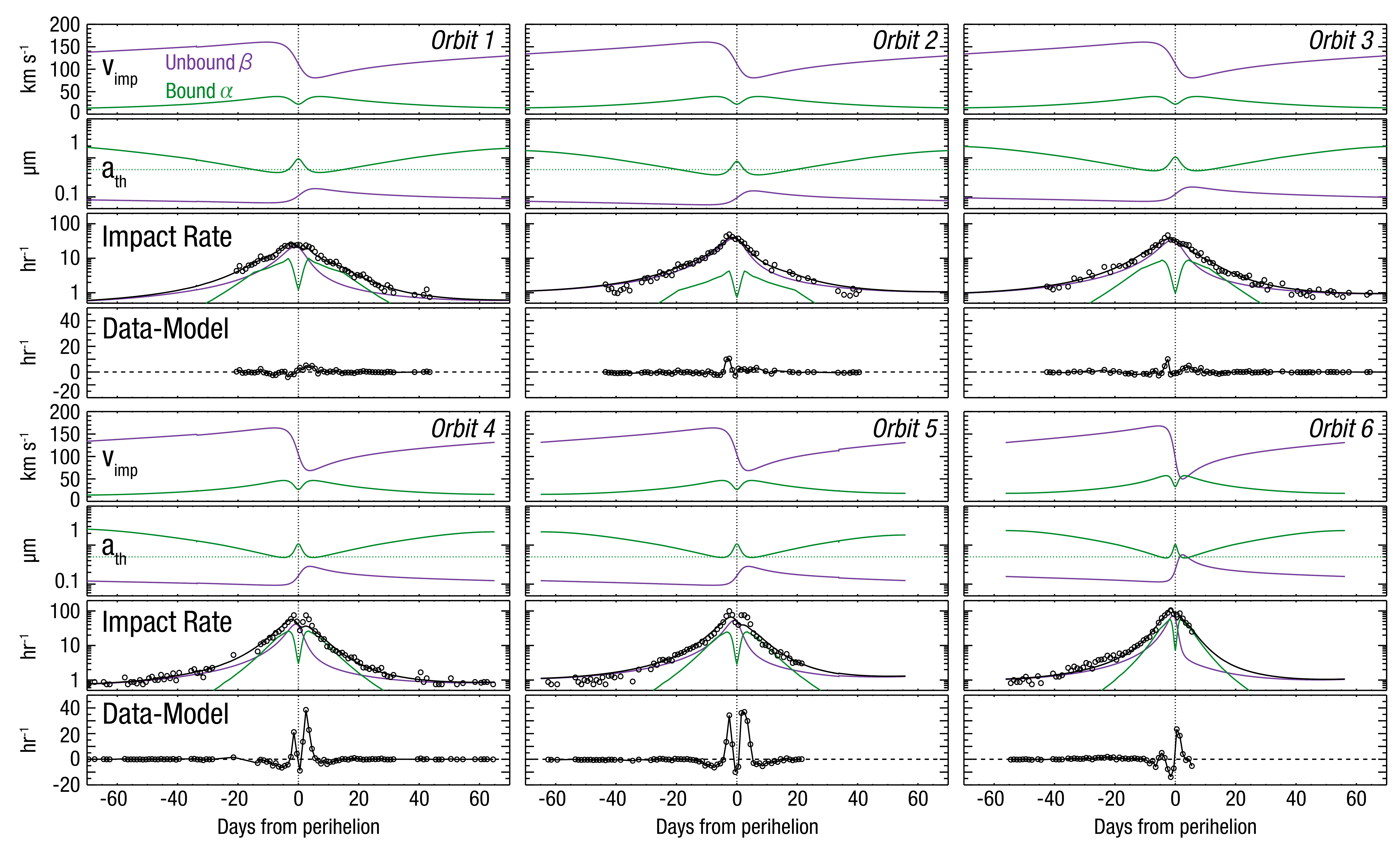}
\caption{Model fits for each orbit. Sub-panels show: modeled impact speed, minimum detectable size, data and model impact rates, and the difference between data and model rates for orbits 1-6. \label{fig:fits}}
\end{figure}

Figures~\ref{fig:radial} and \ref{fig:fits} show the results of the model fits, where each orbit arc is shown separately in Figure~\ref{fig:radial} along with model curves for all orbits up to and including 24 to illustrate the expected profiles to be encountered in future orbits. In Figure~\ref{fig:fits}, the top sub-panels show the expected impact speeds. The 2nd sub-panels show the minimum detectable size threshold corresponding to the fitted {\color{black}$\tilde{Q}_c \equiv Q_c/C$. These values of charge threshold, given in Table~\ref{table:fit}, are similar to our back of the envelope estimate of $Q_c \approx 0.3 \times 10^9$ e$^-$ for $C=0.04 - 0.2$, which is well within the range of expected values for relevant spacecraft surface materials \citep{collette:14b}.} The 3rd sub-panels show the PSP data along with model rates for \amsn, \bmsn, and combined total impact rate. The last subpanels show the difference between the data and modeled rates. 

As perihelion is reduced, each successive orbital group shows an increasing prominence in \ams vs. \bmsn. For orbits 1-3, the large majority of the impact rate profile can be reproduced with \bms, notably for orbit 2 where almost no \ams are found to contribute to the impact rates. For orbits 4 and 5, which have a lower perihelion distance, the post-perihelion portion has a region that is dominanted by \ams. Finally, for orbit 6, both the pre-perihelion and post-perihelion arcs are dominanted by \ams near close approach.

Our model is able to reproduce the overall structure of the PSP measured impact rates for each orbit with three notable exceptions. The first is outside $\sim$0.7 au, where the model predicts larger counts than observed by PSP. This feature is likely due to the non-standard orientation of the spacecraft due to communications attitude changes via yaw maneuvers at the outermost portions of the orbit. For example, the lower counts in Orbit 2's inbound arc before and after $\sim$0.75 au exhibit a step-like change. This change exactly coincides with when PSP's attitude transitioned from a non-standard orientation with the heat shield off-pointed from the solar direction to one in which PSP's heat shield was directly pointed to the solar direction. We use an effective collecting area assuming the PSP +X vector is always aligned with the ram direction and that the +Y vector is aligned with PSP's orbital plane normal vector, which is often not the case outside $\sim$0.7 au (see \citet{malaspina:16a} for spacecraft coordinate definitions). Outside $\sim$0.7 au, PSP is often oriented with rotations about multiple axes compared to its nominal solar pointed ram configuration. Estimating the effective area more precisely for these configurations would require time-dependent 3D modeling of the spacecraft geometry and could be investigated in future studies.

The second feature not predicted by the model is the magnitude of the pre-perihelion peak impact rates. A pre-perihelion peak in impact rate is expected for \bmm impactors due to the enhanced impact speed on the inbound orbit arc \citep{szalay:20a}. Since the impact rate from \ams is symmetric about perihelion, this leads to modeled rates that always peak before perihelion. While the model predicts such a feature and is able to reasonably reproduce the location of this peak, it is not able to fully reproduce the large impact rates observed by PSP. Specifically, orbits 2, 4, and 5 exhibit pre-perihelion enhancements over the model fits of approximately {\color{black}10 hr$^{-1}$, 20 hr$^{-1}$, and 35 hr$^{-1}$} respectively.

The final aspect for which the model deviates from the data is in the post-perihelion rates within $\sim$5 days from perihelion. Across all six orbits discussed here, the model is deficient in reproducing larger impact rates in this region. During orbits 1-3, and 6, this enhancement manifests itself as an extended ``shoulder'' on the impact rate profile, which still monotonically decreases from the expected pre-perihelion peak. During orbits 4-5, this enhancement forms a second peak in the total observed impact rates, with the data-model difference also clearly showing this unexpected feature. The difference between data and model rates for the post-perihelion enhancement maximizes during orbits 4-5. No such feature is predicted by the two-component model and we present possible explanations for this post-perihelion enhancement in Section~\ref{sec:enhance}.

\begin{figure}[ht!]
\plotone{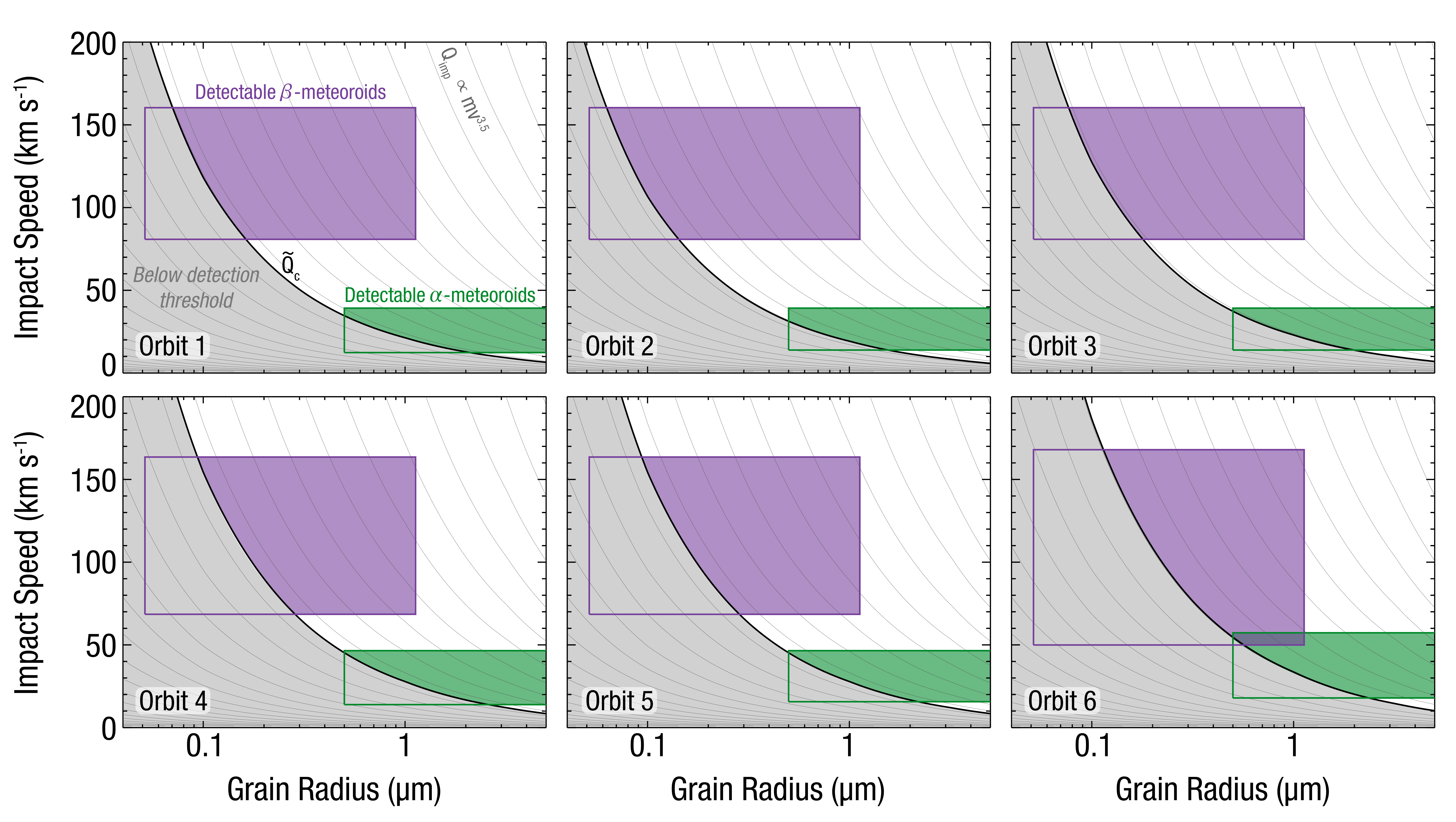}
\caption{Impact speed and size ranges for the two-component model. Grey lines show iso-impact charges following $Q_{imp} \propto mv^{3.5}$, with model fitted detection thresholds ({\color{black}$\tilde{Q}_c$}) shown with the solid line in each sub-panel. Green and purple boxes show the simulated ranges of \ams and \bms respectively, with the shaded portion indicating the detectable range. \label{fig:qplot}}
\end{figure}

The model also allows for a comparison with the fitted impact threshold across the orbits. Figure~\ref{fig:qplot} shows the impact speed and size ranges for the two dust populations in this model. Gray contours show iso-impact-charge features, where any impact on a specific contour is expected to generate the same impact charge and therefore the same impact voltage spike magnitude. The fitted {\color{black}$\tilde{Q}_c$} values are shown in black and the modeled \amm and \bmm populations are marked with the two boxes. The shaded portions of each box shows the range of detectable impactors for each population according to the model fits. As shown here, for the similar values of {\color{black}$\tilde{Q}_c$}, PSP is expected to be able to detect a large portion of available impactor populations. It also shows that even with the difference in both speed and mass of the two populations, the impact charge relation suggests both populations are able to be detected each orbit contemporaneously.

\subsection{Collisional Production Rate in the Zodiacal Cloud}

\begin{figure}[ht!]
\plotone{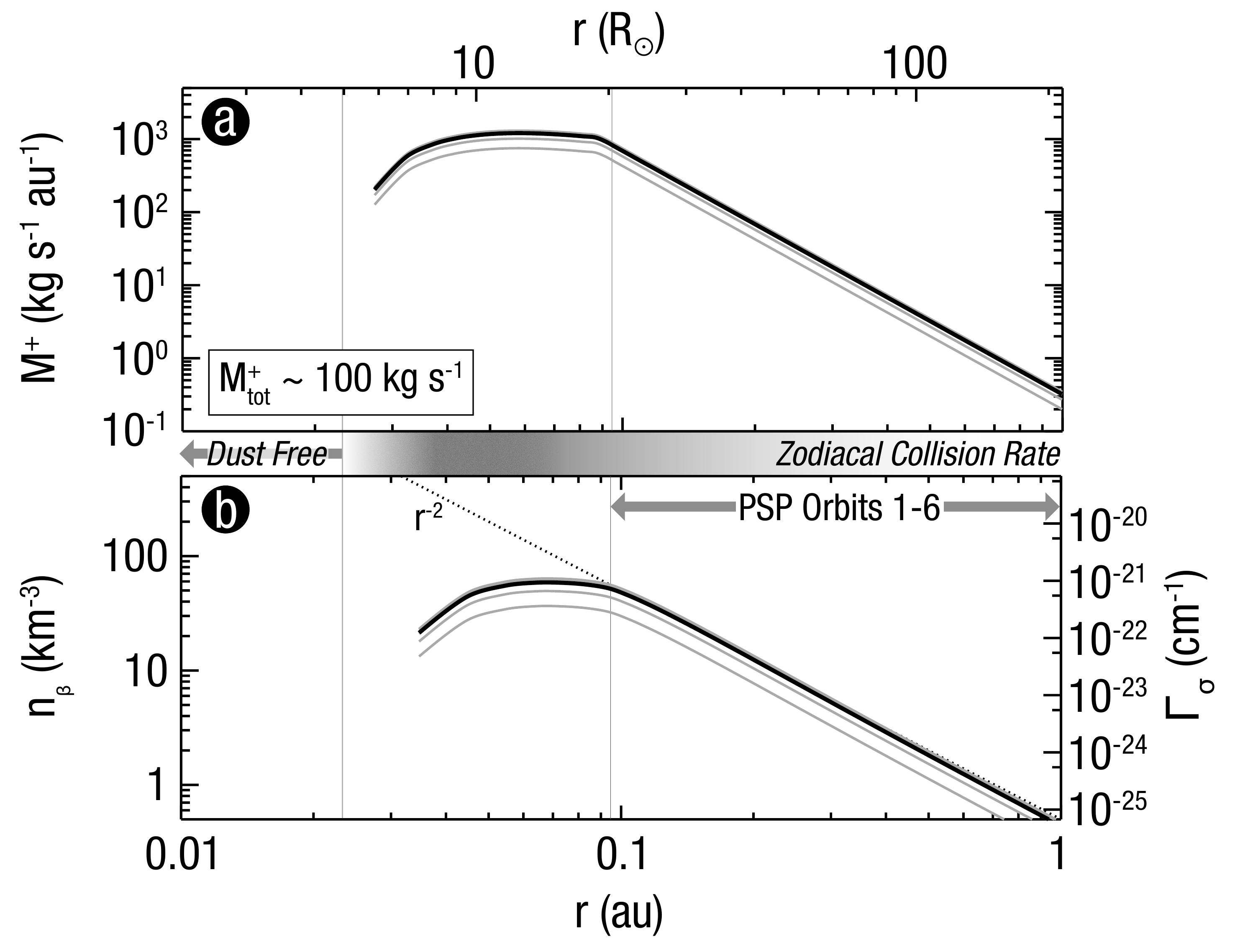}
\caption{Collisional production rate in the zodiacal cloud. a) Forward-modeled production rate per radial distance of \bms consistent with the observed impact rates to PSP. The majority of zodiacal collisions producing \bms occur within approximately 10 to 20 solar radii. b) \bmm density derived from the PSP impact fluxes and forward model. Quantities are displayed as a function of distance from the center of the Sun in au (bottom) and solar radii (top). \label{fig:mdot}}
\end{figure}

Given the model fits to the zodiacal dust cloud, we can provide estimates on the total collisional production rate of grains in the inner solar system. Volumetric mass loss rates are initially calculated in the model assuming \bmm fluxes are uniformly distributed throughout all inclinations as the model is inherently one dimensional. The conversion from observed flux back to $M^+$ assumes a full-sky averaged flux as an input. In reality, the zodiacal cloud has an inclination distribution that consists primarily of low-inclination orbits \citep[e.g.][]{nesvorny:10a}. Since PSP observed the flux near the ecliptic plane, which is larger than the latitudinally averaged flux, we must apply a correction factor. Using a latitudinal dependence \citep{leinert:78b} proportional to $f(\lambda) = 1-\sin{|\lambda|}$, we scale the model mass production values by $\langle f(\lambda)\rangle = 0.36$. {\color{black}Table~\ref{table:fit} shows the results of the fitted total mass production rates for fits of each orbit separately, and combined across all of the first six orbits in the last row.}

The model predicts that the majority of collisions producing \bms that stream throughout the solar system occur within 10 to 20 solar radii (Figure \ref{fig:mdot}a). Inside of $\sim$10 solar radii, the modeled \bmm density monotonically decreases toward the Sun (Figure \ref{fig:mdot}b), indicating a precipitous drop in collisional production near the Sun and suggestive that the size distribution of zodiacal material changes as a function of radial distance \citep{ishimoto:98a}. The location of peak production is consistent with a collisional source mechanism eroding the zodiacal cloud, as opposed to sublimation driven erosion where grains are expected to sublimate inside $\sim$10 solar radii \citep{mann:04a}.  Across all orbits, we find a zodiacal collisional production rate of at least {\color{black}$M^+ = 100$} \kgs \ is consistent with the fluxes of \bms observed by PSP. We have assumed that all collisional products are distributed into \bmsn, hence this number is likely an underestimate, which may explain why it is approximately 5 to 10 times lower than previous collisional estimates \citep{grun:85a}.  

If sub-micron sized grains are responsible for the abundance of pickup ions near the Sun \citep{schwadron:00a}, we expect pickup ion production would be enhanced near the peak \bmm density location around 10 to 20 solar radii.  Previous estimates found a lower limit for the dust geometric cross-section of $\Gamma_\sigma \geq 1.3\times 10^{-17}$ cm$^{-1}$ is needed to produce inner source pickup ions \citep{schwadron:00a}, almost 2 decades larger than the value used previously \citep{holzer:77a,fahr:81a,gruntman:96a}, $\Gamma_\sigma = 2 \times 10^{-19}$ cm$^{-1}$. A geometric cross section for inner source pickup ions of $\Gamma_\sigma \geq 1.3\times 10^{-17}$ cm$^{-1}$ is 4 decades larger than the \bmm geometric cross section found here based on collisional production in the zodiacal cloud; this discrepancy raises the question about whether dust from zodiacal can explain the inner source. 

The distribution of inner source pickup ions indicates that the grain population producing the inner source peaks at $\sim$10 \rst, which is within the region of maximum \bmm collisional production between 10 and 20 \rst \ found here. While not modeled here, collisional fragmentation can also produce grains with radii $\lesssim 50$ nm. This population of nanograins, which are highly susceptible to electromagnetic forces, can become trapped in the inner-most regions of the solar system inside $\sim$30 \rst \citep{czechowski:10a}. Impacts from nanograins in the range of $30-40$ nm have been estimated to have impact speeds in the range of $20-45$ \kms \ during PSP's first three orbits \citep{mann:20a}. Such impacts would produce an impact charge of approximately $10^5-10^6$ electrons, {\color{black}orders of magnitude lower than the estimated detection threshold of $\tilde{Q}_c \sim 10^9$ electrons even accounting for a few orders of magnitude variation in $C$ \citep{collette:14b}.} Hence, this population of nanograins is not expected to be directly detectable in the FIELDS impact data. 

Dust grains can serve as the neutralizing agent for solar wind ions that penetrate these grains, and could potentially explain the large fluxes of inner source pickup ions observed \citep{schwadron:00a}. \citet{wimmer:03a} suggest that the size of such dust particles is less than or comparable to the penetration range of solar wind ions in dust material, $<$100 nm. Thus, the collisional production of zodiacal grains discussed here also suggests the generation of extremely small dust particles that may explain the origin of inner source pickup observed in the inner heliosphere \citep{schwadron:00a}. We have ruled out larger \bms for the inner source of pickup ions, given their low geometric cross section due in part to their relatively short lifetimes in the inner solar system. If dust is responsible for generating the inner source, then it must be from nanograins with radii below $\sim$50 nm.  While this nanograin population is likely not detectable directly with FIELDS measurements, PSP may be able to observe pickup ions in the inner solar system directly with PSP's SWEAP instrument \citep{kasper:16a}. These pickup ions can also seed energetic particles accelerated via shocks in the inner solar system that could be detected by the \isois instrument \citep{mccomas:16a,mccomas:19a}. Hence, the inner-most PSP orbits may yield critical information on the inner-source of pickup ions across its in-situ instrument suites.

\section{Post-perihelion enhancement} \label{sec:enhance}

A post-perihelion impact rate enhancement is not predicted by the two-component model and is observed in every orbit in varying magnitudes. Since PSP cannot directly measure dust density and speed distributions, we use a model to estimate these quantities from count rates. The enhancement in the peak rates above the two-component model rates in orbits 4-5 amounts to a measured flux of approximately $3 \times 10^{-3}$ m$^{-2}$ s$^{-1}$, which correspond to densities in the range $30-300$ km$^{-3}$ for $v_{imp} = 10-100$ \kms. We investigate two possibilities to explain these post-perihelion enhancements, both of which are related to meteoroid streams: direct meteoroid stream encounters or \bms created by collisions between meteoroid streams and the nominal zodiacal cloud.

\subsection{Direct meteoroid stream encounter}

The estimated densities necessary to account for the post-perihelion enhancement are  similar to densities of meteoroid streams inferred by Helios in-situ dust measurements of $10-100$ km$^{-3}$ \citep{kruger:20a}. Applying a similar analysis, we utilize the Interplanetary Meteoroid Environment for eXploration (IMEX) model \citep[e.g.][]{soja:19a} to estimate the number flux PSP encounters near transits with cometary orbits. This model tracks the trajectories of grains with radii greater than 100 $\mu$m that are released due to cometary activity to build up a density distribution along the comets orbit. It accounts for gravitational and solar radiation forces acting on dust grains, as well as the collisional lifetimes. Grains with radii smaller than 100 $\mu$m are also likely to exist in a more extended cross-sectional area, however, they dynamically decouple from their parent comets' orbits much more rapidly for smaller sizes. 

\begin{figure}[ht!]
\plotone{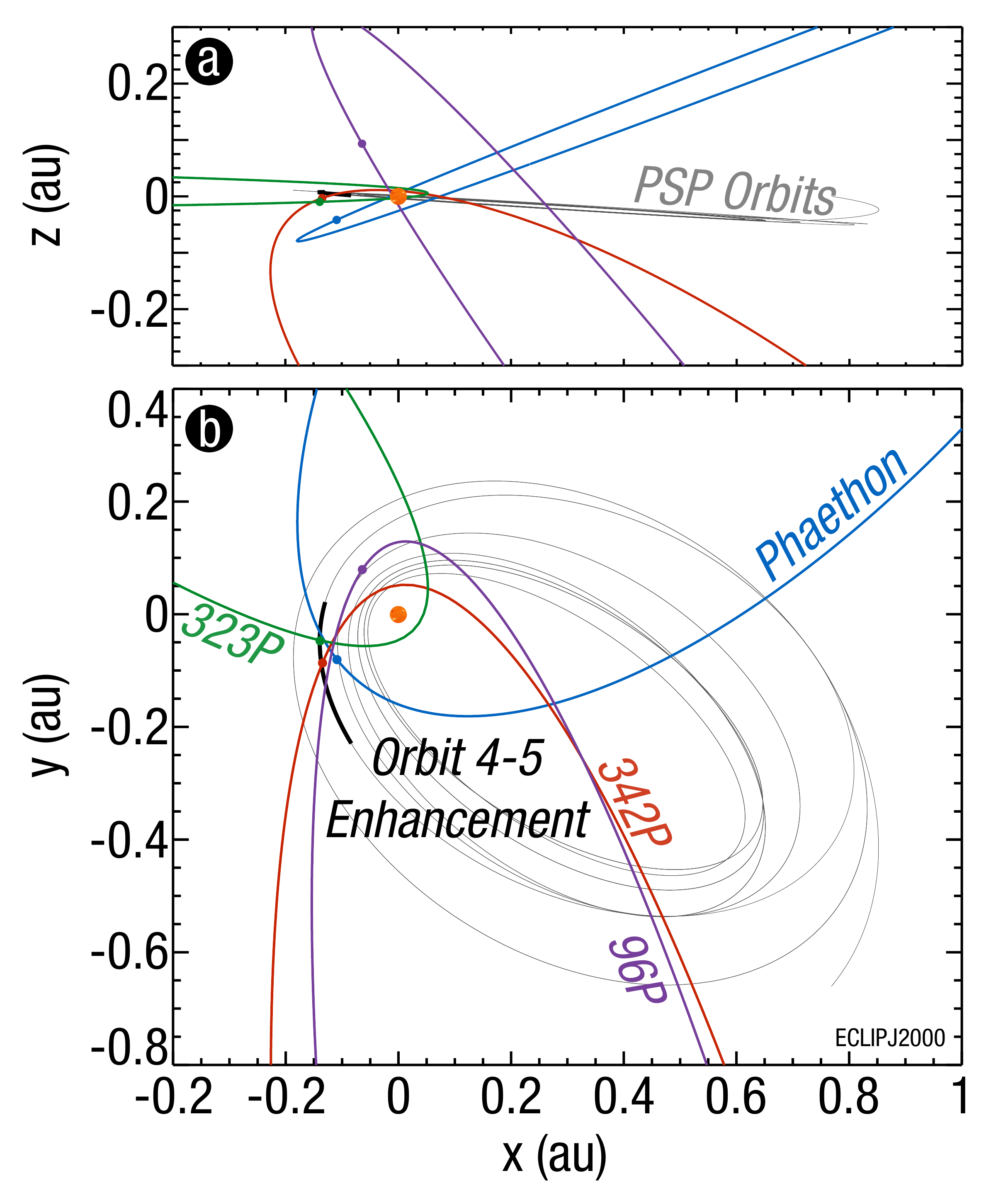}
\caption{Orbits of meteoroid stream parent bodies that transit near PSP's orbit 4-5 post-perihelion impact rate enhancement in the ECLIPJ2000 frame for side (a) and top-down (b) views. \label{fig:comets}}
\end{figure}

\begin{table}
\begin{centering}
\begin{tabular}{cccc}
Body & $d_\mathrm{min}$ & $v_\mathrm{imp}$ & $n_\mathrm{lim}$ \\
 & (au) (\rst) & (km s$^{-1}$) & (km$^{-3}$) \\
\hline \hline
(3200) Phaethon & 0.056 (12) & 62 & 48 \\
342P/SOHO & 0.0069 (1.5) & 54 & 55 \\
323P/SOHO & 0.017 (3.7) & 131 & 23 \\
96P/Machholz & 0.12 (25) & 87 & 34\\
\hline \hline
\end{tabular}
\caption{\label{table:comets}Meteoroid stream candidates. At the time of closest approach of each orbit to PSP's orbit, $d_\mathrm{min}$ is the minimum distance, $v_\mathrm{imp}$ is the impact speed, and $n_\mathrm{lim}$ is the density limit that would be needed to generate an impact flux of $3 \times 10^{-3}$ m$^{-2}$ s$^{-1}$ at PSP.}
\end{centering}
\end{table}

We performed a comprehensive search across all known comets in the inner solar system to determine which orbits have the closest approach points near the region of peak impact rate enhancement in PSP's orbits $4-5$. From this search, we identified three cometary candidates: 342P/SOHO, 323P/SOHO, and 96P/Machholz (Table~\ref{table:comets}). We also investigate the Geminids meteoroid stream, whose parent body is likely asteroid (3200) Phaethon. Figure~\ref{fig:comets} shows these four orbits along with the PSP orbits and region of enhanced post-perihelion impact rates for orbits 4-5. Dots mark the location on each orbit closest to the Orbit $4-5$ enhancement. As shown in this figure, the stream parent bodies identified have close transits with PSP's orbit, with 342P transiting the nearest to the region of interest. For the cometary candidates, only one had sufficient brightness data to constrain the stream number densities: 96P. For this stream, we find its impact rates would peak a few days before the observed enhancement with an upper limit of $\sim$10$^{-3}$ hr$^{-1}$, making it an unlikely explanation for the post-perihelion enhancement. 

Estimating the peak rates for 342P and 323P requires comprehensive knowledge of the absolute brightness of these comets not available at the time of this analysis. Of these two candidates, 323P/SOHO does not exhibit activity like a normal comet \citep{battams:17a}; its activity is similar to 322P/SOHO which exhibits activity more similar to an asteroid than a comet \citep{knight:16a}. Thus, dust production from 323P is likely minimal, and a substantial debris trail is unlikely. While we cannot directly estimate these meteoroid stream densities, with knowledge of the impact speeds from each stream, we can estimate the required densities necessary to produce the orbit 4-5 enhancement, $n_\mathrm{lim}$ given in Table~\ref{table:comets}. These density limits are at the position of PSP. Given that PSP flies within a few solar radii of 342P and 323P, these density limits are applicable to core of the stream. Note, $n_\mathrm{lim}$ is the limiting density needed to produce the observed impact rates; it does not signify that any of these meteoroid streams actually have these densities, as that would require an exact determination of the source of the PSP impact rate enhancement.

Phaethon, the parent body of the Geminids meteoroid stream, does not continually replenish this stream in sufficient quantities via the relatively low activity level of active asteroid Phaethon \citep[e.g.][]{jewitt:10a,jewitt:13a}, nor is meteoroid bombardment of Phaethon sufficient to source the stream \citep{szalay:19a}. However, unlike the other meteoroid streams discussed here, we have direct observations of its flux. At 1 au, the number flux of Geminids with radii $\geq$100 $\mu$m is $4 \times 10^{-11}$ m$^{-2}$ s$^{-1}$, which corresponds to a number density of $1.2 \times 10^{-6}$ km$^{-3}$ for their heliocentric speed of 33 \kms \  \citep{blaauw:17a}. Its closest encounter with the PSP trajectory is at $\sim$0.14 au, where the Geminids have a heliocentric speed of $\sim$109 \kms. Scaling the density as $r^{-1.5}$ (Appendix~\ref{app:rate}) at 0.14 au gives an expected density of $2 \times 10^{-5}$ km$^{-3}$. With a relative impact speed between PSP and the Geminids of 62 \kms, an effective area of 6 m$^2$, and assuming PSP flew directly through the core of this meteoroid stream, PSP could expect impact rates of $3 \times 10^{-5}$ hr$^{-1}$ for grains with radii $\geq 100$ $\mu$m, well below the rates necessary to explain the impact rate feature. The cumulative mass index of the Geminids was found to be $ \alpha_g = 0.68$. Even if we assume a power-law size distribution holds to radii as small as 1 $\mu$m, the expected impact rate from grains $\geq$1 $\mu$m would be 0.4 hr$^{-1}$, still significantly lower than the {\color{black}$\sim$40 hr$^{-1}$} observed in the enhancement. PSP also does not transit directly through the core of the stream, its close approach point during the enhancement period is approximately 0.06 au from the core of the stream. Therefore, we do not attribute the enhancement to direct Geminids observations, however, we explore another alternative related to the Geminids in the next section.

\subsection{\bs encounter}

\begin{figure}[ht!]
\plotone{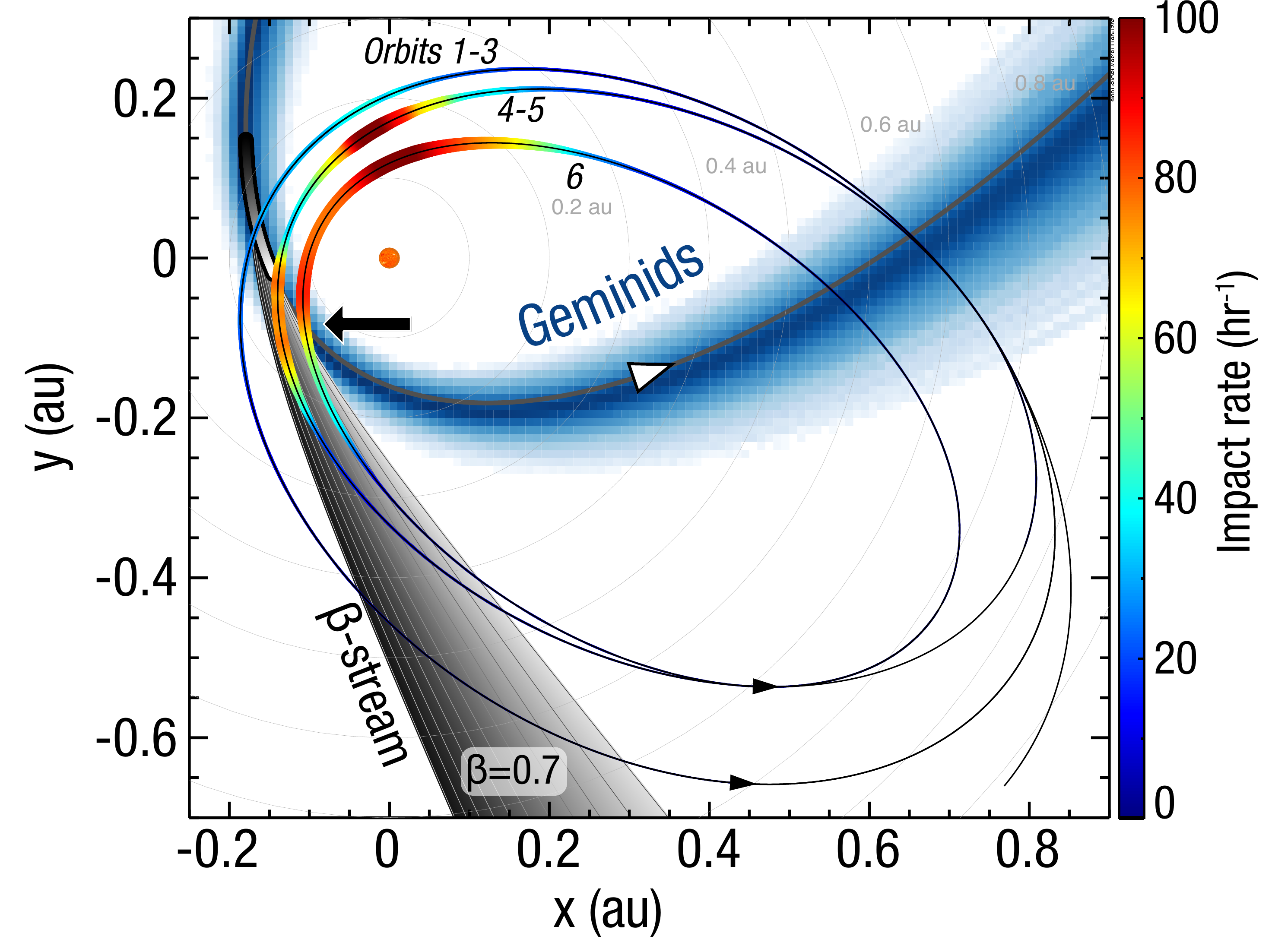}
\caption{\bs geometry. Impact rates overlaid on the PSP trajectory in the ecliptic J2000 frame, averaged over orbits 1-3 for the outer-most orbit and individually shown for orbits 4-6. Color and width of the color strip encode rate. The gray fan-like shape shows the example path taken by \bms for $\beta=0.7$ produced in region in the Geminids that accounts for 95\% of detectable fluxes during PSP's orbits 4-5. Blue contours show the approximate density profile of larger meteoroids in the Geminids stream. 
}
\label{fig:gem}
\end{figure}

As an alternative to a direct meteoroid stream encounter, the post-perihelion enhancement could be due to \bms produced by collisions between meteoroid streams and the nominal zodiacal dust distribution. The timing and location of the post-perihelion enhancement coincide with PSP’s intersection of \bmm trajectories produced by collisions between the Geminids meteoroid stream and the nominal zodiacal cloud (Fig. \ref{fig:gem}), which we term a ``\bsn''. From previous estimates of the Geminids collisional lifetime and total stream mass, the stream losses $\sim$6 to 40 \kgs \ from collisions and would register impact rates on PSP in the range of $\sim$0.1 to 30 hr$^{-1}$ during the post-perihelion enhancement period for orbits $4-5$ (Appendix~\ref{app:rate}). The enhanced impact rate features post-perihelion above the modeled rates are therefore qualitatively consistent with Geminids \bs fluxes. Figure~\ref{fig:gem} shows the possible detection geometry for a Geminids \bsn. The blue contours show the approximate density profile of larger meteoroids in the Geminids stream, integrated from $>$14,000 orbits of visual meteors \citep{jenniskens:18a}. The gray fan-like shape shows the example path taken by \bms for $\beta=0.7$ produced in the region in the Geminids (grey tube) that would account for 95\% of potentially detectable fluxes during PSP's orbits $4-5$.

A \bs detection could also explain why PSP observed a post-perihelion enhancement for each orbit, as the \bs would be expected to be dispersed and could intersect multiple orbits. However, to reach PSP from the core of the Geminids stream, the collisional process must generate fragments with transverse spread $\sim$10-20$^\circ$ out of plane above their original trajectories (Figure~\ref{fig:gem_side}). If the enhancement is due to the Geminids \bsn, it suggests the Geminids stream could be more spatially extended than inferred from the core of larger grains which cause visual meteors at Earth (Appendix~\ref{app:geo}).

There are many additional meteoroid streams which could produce \bss detectable by PSP, listed in the Appendix in Table~\ref{table:bs}. {\color{black}The Geminid \bs is preliminarily a favorable candidate and initial analyses on the directionality inferred by FIELDS observations also support this source mechanism \citep{pusack:21a}. A  comprehensive comparison between multiple meteoroid stream origins along with directionality and amplitude analysis} would be crucial in determining the source of the post-perihelion enhancement.

\section{Implications for PSP's Inner Orbits} \label{sec:predict}
\subsection{Predicted impact rates}
\begin{figure}[ht!]
\plotone{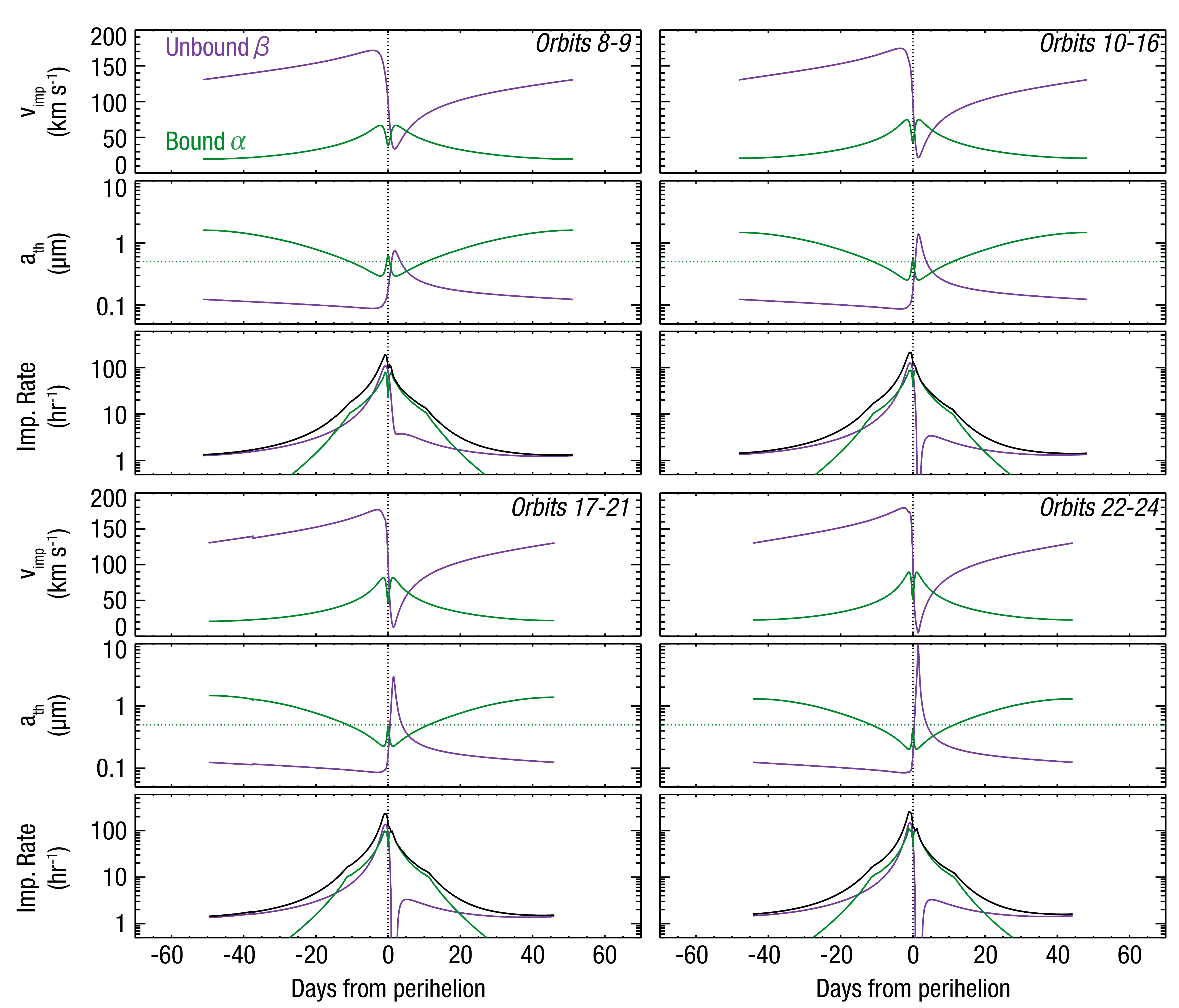}
\caption{Model predictions for PSP orbits 8-24, with sub-panels described in Figure~\ref{fig:fits}. \label{fig:predict}}
\end{figure}

After orbit 7, PSP's perihelion is inside $r_1 = 19 \rs$ (Table~\ref{table:predict}), such that the density of bound \ams on elliptic orbits is no longer expected to increase with a power-law relation. Therefore, inside this critical distance, the PSP impact rates will be more governed by the relative impact speed to dust. Figure~\ref{fig:predict} shows the expected impactor parameters for the two-component model in the upcoming orbital groupings, along with the peak impact speeds and rates for \ams and \bmsn. Figure~\ref{fig:radial} also highlights the expected impact rates for subsequent orbits in the four grey curves with the largest impact rates.

\subsection{Deposited Impact Energy to PSP}
Dust presents a hazard to the PSP spacecraft, both via direct impacts to instruments and to the spacecraft subsystems. On orbit 2, the \isois instrument registered a permanent increase of noise in one of its 80 aperatures which has been attributed to a dust impact \citep{hill:20a}. The timing and orientation of the affected aperture during the noise event suggested the impact was a \bmm \citep{szalay:20a}, as it occurred exactly when the flux of \bms to that aperture was expected to peak and the orientation at that time would have prevented \ams on perfectly circular orbits from impacting the aperture foil. Initial models suggested a total of 1-10 more impacts of this type could occur between orbits 1-6 \citep{szalay:20a}, however, no more apertures on \isois have registered a similar permanent noise-inducing event. Additionally, the WISPR instrument often records images which are dominated by features attributed to impact ejecta spray from the spacecraft due to dust impacts. 

\begin{table}[ht!]
\begin{centering}
{\color{black}\begin{tabular}{ccccccccccc}
Orbit & $r_\textrm{ph}$ & $v^\textrm{max}_{\alpha}$ &  $v^\textrm{max}_{\beta}$ & $R_\alpha$ & $R_\beta$ & $R_{tot}$ & $\frac{E_\alpha}{E_\alpha(6)}$ & $\frac{E_\beta}{E_\beta(6)}$ & $\frac{EF^\textrm{max}_\alpha}{EF^\textrm{max}_\alpha(6)}$  & $\frac{EF^\textrm{max}_\beta}{EF^\textrm{max}_\beta(6)}$ \\
 & (au) (\rst) & (km s$^{-1}$) & (km s$^{-1}$) & (hr$^{-1}$)  &  (hr$^{-1}$)  & (hr$^{-1}$) \\
\hline \hline
1-3 & 0.16 (36) & 39 & 160 & 6 & 31 & 33 & 0.26 & 0.62 & 0.16 & 0.37\\
4-5 & 0.13 (28) & 46 & 163 & 20 & 49 & 59 & 0.45 & 0.71 & 0.36 & 0.58\\
6-7 & 0.095 (20) & 57 & 167 & 51 & 83 & 130 & 1 & 1 & 1 & 1\\
8-9 & 0.074 (16) & 66 & 171 & 79 & 110 & 187 & 1.7 & 1.2 & 2.2 & 1.5 \\
10-16 & 0.062 (13) & 74 & 174 & 88 & 120 & 212 & 2.2 & 1.3 & 3.7 & 1.9\\
17-21 & 0.053 (11) & 81 & 176 & 97 & 140 & 228 & 3.0 & 1.5 & 5.3 & 2.3 \\
22-24 & 0.046 (9.9) & 89 & 179 & 110 & 150 & 253 & 3.7 & 1.6 & 7.3 & 2.7 \\
\hline \hline
\end{tabular}}
\caption{\label{table:predict} Predicted quantities during each of PSP's orbital groups. $r_\textrm{ph}$ $-$ perihelion distance, $v^\textrm{max}$ $-$ peak impact speed, $R$ $-$ peak detectable impact rate, $E$ $-$ total deposited energy throughout the orbit, and $EF^\textrm{max}$ $-$ peak energy flux. Ratios are given with respect to values during orbit 6.}
\end{centering}
\end{table}

Table~\ref{table:predict} gives an indication of the expected maximum impact speed and rate or all orbit groups. By orbit 6, the \bmm impactor population has a peak impact speed of 167 \kms, not far from its overall mission peak of 179 \kms. Between orbit 6 and 24, the impact rate of \bms is expected to approximately double. Therefore, the \bmm population is expected to minimally increase in overall impact energy for the remainder of the mission. On the other hand, the \amm population increases significantly in both impact speed and total impact flux. We estimate the peak and total deposited energy for each orbit. The peak deposited energy flux is determined by multiplying the total peak impact rate (independent of detection threshold) by the average  energy per impact. The total deposited energy is the integral of the impact rate multiplied by average energy per impact along the entire orbit. Table~\ref{table:predict} lists these quantities as ratios to their values during orbit 6 to give an indication of how much these quantities increase in subsequent orbits. As given in this table, the peak energy flux is expected to be as much as $\sim$7 times larger during orbits $22-24$ compared to orbit 6. Additionally, the possibility exists that PSP could directly encounter meteoroid streams from cometary activity, posing a hazard in addition to those discussed here.

\section{Discussion and Conclusions} \label{sec:discuss}
PSP impact rates carry critical information on the collisional environment in the inner solar system. Initial analyses of the first three PSP orbits concluded the impact rates were consistent with \bms \citep{szalay:20a,page:20a,malaspina:20a}, with a subsequent analysis finding electromagnetic forces were mostly negligible for the dynamics of the observed impactor population \citep{mann:20a}. With six orbits of data spanning three separate orbital families, the impact rate profiles have revealed additional substructure that is not explainable by \bms alone, particularly after orbit 3. Given that PSP does not have a dedicated dust detector capable of measuring the mass and/or velocity distributions of impacting grains, a number of assumptions must be made to interpret this dataset, simplify the data-model comparison framework, and minimize the number of free modeling parameters. We find that the broad, overall impact rate profiles are consistent with two dust components comprised of bound \ams on elliptical orbits and unbound \bms on hyperbolic orbits; additional more localized sources might account for smaller-scale features.

\begin{figure}[ht!]
\plotone{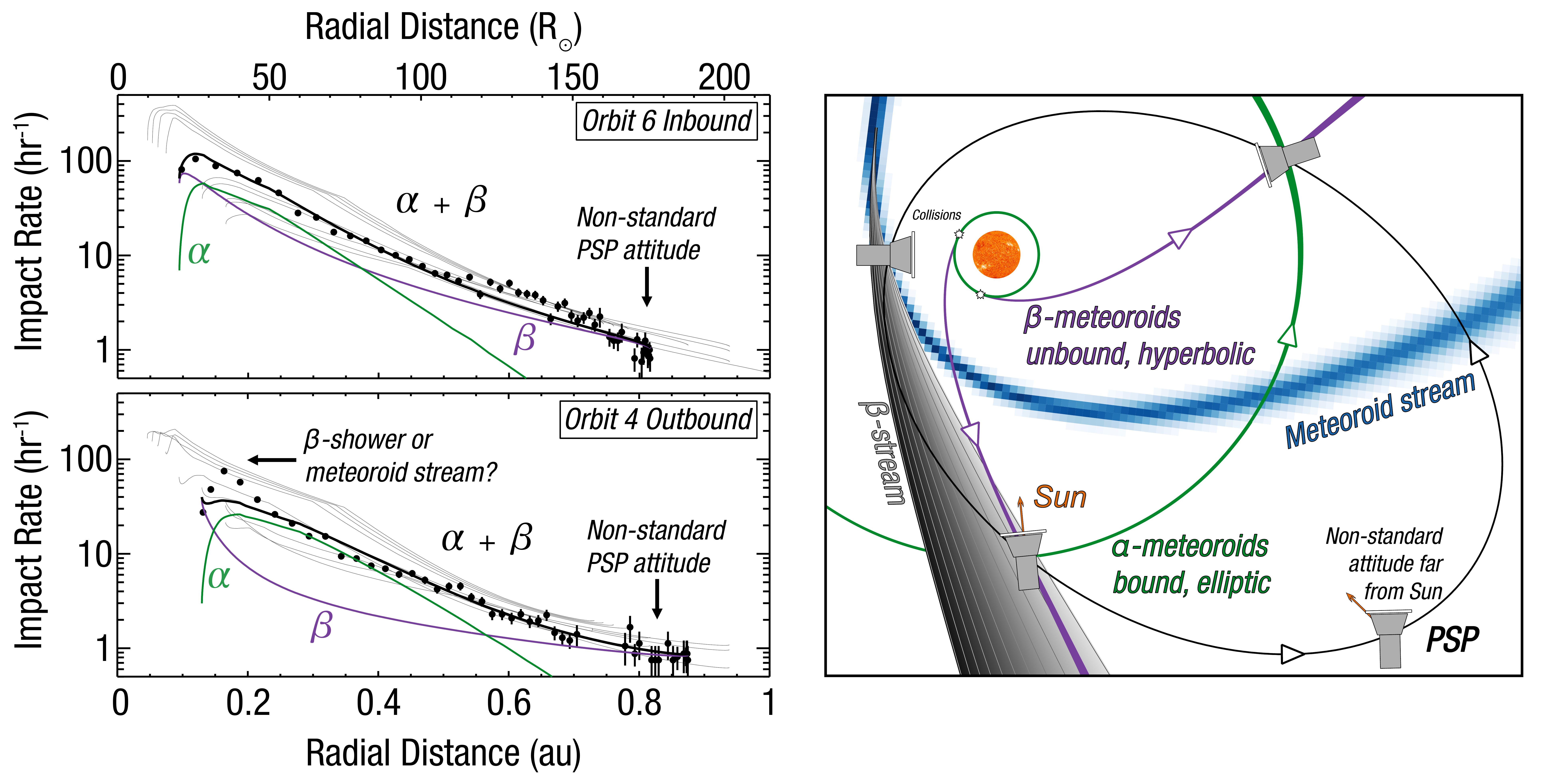}
\caption{Summary showing the various identified and potentially detected dust populations along with data from orbit 6 (inbound) and orbit 4 (outbound) to highlight characteristic features observed across orbits 1-6. \label{fig:summary}}
\end{figure}

Figure~\ref{fig:summary} highlights the various dust populations discussed in this work, as well as the three notable features in the data that the model is not able to fully reproduce. First, outside $\sim$0.7 au, the model predicts larger impact rates than observed due to non-standard and rapidly changing spacecraft orientations which reduce PSP's effective area. Second, while the model predicts a pre-perihelion impact rate peak for each orbit due to the asymmetric impact speeds for \bms during the inbound and outbound portions of each PSP orbit, it is not able to fully reproduce the magnitude of this enhancement for a subset of the first six orbits. A comprehensive dynamical model \citep[e.g.][]{pokorny:19b} could potentially reveal additional substructure not captured due to the assumptions involved in the idealized two-component model discussed here and may better predict this pre-perihelion enhancement. We note the impact rate due to grains with radii $\sim$10's nm that experience non-negligible electromagnetic forces are expected to be even more asymmetric than those considered here \citep{mann:20a}. Such a population, which was not modeled here, is expected to sharply peak in impact rate before perihelion. However, such a population is not expected to be directly detectable, as it would generate impact charges orders of magnitude lower than our estimated impact threshold.

The largest discrepancy between the data and model occurs a few days after perihelion. All six orbits' impact rate profiles exhibit an enhancement above the model predictions during this period, most evident in the sharp second peaks during orbits 4 \& 5. We suggest two possible explanations originating from a meteoroid stream: 1) direct encounters with meteoroid streams or 2) \bssn, the collisional byproducts of meteoroid streams. A \bs from the Geminids meteoroid stream is a favorable candidate and estimates suggest such a process could provide the necessary fluxes to explain this feature. It is also favorable as such a population could potentially be observable across multiple of PSP's orbital families. However, determining the origin for this impactor population presents difficulties, particularly given that FIELDS is not a dedicated dust instrument. Additional information on the directionality and amplitude distribution {\color{black}has provided additional critical insight specifically on the orbit 4 enhancement and preliminarily favors the Geminids \bs origin \citep{pusack:21a}; a comprehensive comparison of directionality and amplitude distributions across multiple orbits would help further characterize the origin of this third impactor population.}

If the post-perihelion enhancement is due to a Geminids \bsn, it would be the first direct observation that asteroidal and cometary debris trails collisionally erode as they transit the zodiacal cloud. Hence, PSP may be an efficient laboratory to investigate collisional processes and could monitor the activity and catastrophic breakup of asteroids and comets in the inner solar system. It could also reveal additional meteoroid streams, either directly or through their \bssn, that would be difficult to detect via other means; such a detection would be similar to the observation of an intense meteoroid shower at Mercury attributed to Comet Encke, which does not intersect Earth's orbit \citep{christou:15a}.  {\color{black}Many of the candidate parent bodies for the enhancement have low activity. If the enhancement is due to one these streams, it could indicate additional streams similar to the Geminids exist, where the total stream mass is comparable to the mass of the parent body likely as the result of a catastrophic event in the past few thousand years.}

By fitting to the observed PSP impact rates, we estimate the total zodiacal collisional mass production rate to be at least {\color{black}$M^+ = 100 $ \kgs, approximately an order of magnitude} lower than previous collisional estimates \citep{grun:85a}. The discrepancy between these two mass estimates is likely related to our assumption that all collisional products become \bmsn, suggesting our estimate is a lower limit. Propagating the model results to 1 au, the flux of \bms  (Table~\ref{table:fit}) is estimated to be {\color{black}$0.4-0.8 \times 10^{-4}$ m$^{-2}$}. This is similar to previous PSP estimates of $0.3-0.7 \times 10^{-4}$ m$^{-2}$ based on a simpler \bmm model \citep{szalay:20a} and $0.4-0.5 \times 10^{-4}$ m$^{-2}$ from a distributed \bmm source model for grains with radii between 100 to 140 nm \citep{mann:20a}. These values are notably similar to fluxes of \bms derived from STEREO antenna measurements of $0.2-1 \times 10^{-4}$ m$^{-2}$ \citep{zaslavsky:12a}. All antenna-based estimates are lower than fluxes derived from dedicated dust detectors onboard Pioneers 8 and 9 of $6 \times 10^{-4}$ m$^{-2}$ \citep{berg:73a} and Ulysses of $2 \times 10^{-4}$ m$^{-2}$ \citep{wehry:99a,wehry:04a}. Discrepancies could be due to the different detection mechanisms and relative sensitivies of the various detection mechanisms, or might indicate an intrinsic variability in \bmm flux as suggested from Ulysses measurements \citep{wehry:04a}.

We expect the majority of collisions in the zodiacal cloud occur in the region spanning approximately 10 to 20\rst \ and if the inner source of pickup ions is of dust origin \citep{schwadron:00a}, it must be from grains with radii less than $\sim$50 nm. These findings underscore the prevalence of collisions and \bms within our solar system, which are a crucial part of the lifecycle of the zodiacal cloud. \bms could also be important in space-weathering processes on airless bodies, particularly in the inner solar system where impact speeds and fluxes are high. An airless body with an eccentric orbit like PSP's would experience orders of magnitude higher fluxes and impact speeds for both \ams and \bms compared the Moon \citep[e.g.][]{szalay:19a}, where \bms have been suggested to be an important driver in producing impact ejecta \citep{szalay:20b}. Based on the updated fluxes derived in this work, $\sim$10$^9$ \bms impact the Moon's surface each second. If even a small fraction of those impactors are able to liberate lunar surface material, it could be an important space weathering driver. 

PSP's subsequent orbits will allow us to further probe  the diverse dust populations in the innermost regions of our solar system and enable us to directly compare impact rate measurements with the remote sensing observations \citep{stenborg:20a} that have suggested a departure from the nominal power-law scaling of the zodiacal dust density inside 19\rst. We summarize our results below,
\begin{itemize}
    \item The broad PSP impact rate features in orbits $1-6$ can be explained by two dust sources: \ams on bound, elliptic orbits and \bms on unbound, hyperbolic orbits;
    \item A prominent excursion from model predictions after each perihelion suggests meteoroid streams may be observed, either directly or by their collisional byproducts through a ``\bsn'';
    \item \bss are expected to be a universal phenomenon in all exozodiacal disks;
    \item PSP is likely unable to detect nanograins with radii less than $\sim$50 nm;
    \item Future PSP orbits are expected to experience increasingly intense fluxes from \amsn;
    \item A zodiacal erosion rate of at least {\color{black}$\sim$100} \kgs \ is consistent with the observed impact rates;
    \item The flux of \bms at 1 au is estimated to be {\color{black}$0.4-0.8 \times 10^{-4}$ m$^{-2}$ s$^{-1}$};
    \item The majority of zodiacal collisions producing \bms occur in a region from $\sim$10$-20$ \rst; 
    \item If the inner source of pick-up ions is due to dust, it must be from nanograins with radii below $\sim$50 nm.
\end{itemize}

\acknowledgments

We thank the many Parker Solar Probe team members that enabled these observations. We acknowledge NASA Contract NNN06AA01C. P.P. was supported by NASA Solar System Workings award number 80NSSC21K0153. The IMEX Dust Streams in Space model was developed under ESA funding (contract 4000106316/12/NL/AF - IMEX).



\clearpage
\appendix

\section{Geminids \bs Collisional Production Rate} \label{app:rate}
The collisional lifetime of the Geminids meteoroid stream has been previously estimated as $\tau_c$ = $20 - 40 \times 10^3$ yr \citep{steel:86a}. The total mass in the stream is estimated to be $M_{gem} = 2 - 7 \times 10^{13}$ kg \citep{blaauw:17a}. The average mass loss (or collisional production rate) can then be approximated to be $M^+ = \Delta M_{gem}/\Delta t \approx e^{-1} M_{gem}/\tau_c = 6 - 40$ \kgs. The number production rate is related to the mass production rate as
\begin{equation}
N^+ = M^+ \frac{1-\alpha}{\alpha} \frac{m^{-\alpha}_{min} -m^{-\alpha}_{max}}{ m^{1-\alpha}_{max} - m^{1-\alpha}_{min}} ,
\end{equation}
where $\alpha$ is the cumulative mass index. We use a value of $\alpha = 0.9$, measured for collisional ejecta above the Moon \citep{horanyi:15a}, which is comprised of silicate material that is collisionally ground to sub-micron sized via impacts with speeds of $\sim$$10 - 60$ \kms \  \citep{pokorny:19a}. The lunar ejecta environment is a reasonable proxy for the Geminids \bs collisional products, as they are also expected to be silicate and have relative impact speeds with the zodiacal grains $\sim$$30 - 50$ \kms \ near perihelion. The location of the post-perihelion enhancement was found to be consistent with trajectories of grains released from the Geminids with $\beta \gtrsim 0.5$. For a nominal asteroidal grain \citep{wilck:96a}, this corresponds to a mass range of $m_{min} = 4.7 \times 10^{-18}$ kg and $m_{max} = 1.7 \times 10^{-15}$ kg, or grain radius range of 70 to 500 nm. Assuming the entirety of Geminids collisional byproducts are created in this mass range, we estimate $N^+ = 1.7 - 12 \times 10^{17}$ s$^{-1}$ grains are collisionally produced. 

The collisional probability between the Geminids and nominal zodiacal cloud is expected follow $P = n_g n_\alpha v \sigma_{gz} \Delta V$ where $n$ is number density, $v$ is the average impact speed, $\sigma$ is the collisional cross section, $\Delta V$ is the volume the collisions occur in, and subscript `g'  corresponds to the Geminids \citep{steel:86a}. Assuming the cross-sectional area of the Geminids tube is proportional to $r^2$ and the orbital speed is proportional to $r^{-0.5}$, the Geminids density is estimated to follow $n_g(r) = n_{0g}r^{-1.5}$, where $r$ is radial distance. To calculate the impact speed, we assume the Geminids stream always impacts zodiacal grains on circular orbits with the same inclination. The impact speed is 
\begin{equation}
v^2 = \mu \left(  \frac{3}{r} - \frac{1}{a} - 2\sqrt{ \frac{a(1-e^2)}{r^3} } \right),
\end{equation}
where we use the Geminids parent body Phaethon's orbital elements of semi-major axis $a = 1.27$ au and eccentricity $e = 0.89$ as a proxy for the Geminids. The collisional cross sectional area is dominated by the much larger Geminids grains, therefore $\sigma = \pi s_g^2$.

Incorporating the latitudinal dependence \citep{leinert:78b} into the zodiacal density distribution given in Equation~\ref{eq:na}, we assume the cumulative density for grains with radii $>s$ is 
\begin{equation}
n_\alpha(r,\lambda,a) = n_{0\alpha}f_\alpha(r)(1-\sin{|\lambda|})r^{-1.3} \left( \frac{s_\alpha}{s_0} \right)^{-3\alpha},
\end{equation}
where $\lambda$ is the ecliptic latitude. For simplicity, we assume the cumulative mass index $\alpha$ of the zodiacal cloud is the same as the Geminids. The value of $s_\alpha$ is the minimum size zodiacal grain that can catastrophically destroy a Geminids grain upon impact. A key parameter to estimate is the ratio of mass in an impact required to catastrophically destroy the larger mass, $\Gamma = m_g / m_\alpha$, where $m_g$ is the larger mass as the Geminids is comprised of grains in the size range of 100 $\mu$m to 1 mm \citep{blaauw:17a}. $\Gamma$ is related to the threshold energy \citep{grun:85a} $Q_s = v^2/2\Gamma$. Hence, the size of a bound zodiacal grain that can catastrophically destroy a Geminids grain is
\begin{equation}
s_\alpha = \left(   \frac{3Q_s m_g}{2\pi \rho_\alpha u^2}  \right)^{1/3},
\label{eq:sc}
\end{equation}
where $\rho_\alpha$ is the average grain mass density. The last term in $P$, the collision volume, is calculated per constant unit length along the Geminids tube to be $\Delta V \propto r^2$. To determine the region where the majority of collisions occur, we express collision probability as a proportionality relation for all varying terms,
\begin{equation}
P \propto (1-\sin{|\lambda|}) v(r)^{2\alpha+1}r^{-0.8}
\end{equation}

PSP would be impacted by collision products a distance $d$ from the source region. Assuming the \bs is produced in a cone of solid angle $\Omega$, the flux to PSP is
\begin{equation}
F = \frac{N^+}{\Omega d^2} \left(\frac{v_\mathrm{imp}}{v_\mathrm{g}} \right).
\end{equation}
The normalized product $FP$ gives the amount of detectable flux at PSP. Using PSP's location during its peak anomalous post-perihelion enhancement for orbits 4 and 5 as a reference position, 95\% of the flux it could detect from the Geminids \bs originates from a portion in the Geminids tube within 0.15 to 0.27 au pre-perihelion (grey tube along Geminids orbit, Figure~\ref{fig:gem}, when PSP is within 0.09 to 0.29 au from the core of the stream. We estimate this region to account for $\sim$12\% of the total collisional mass loss in the Geminids. A speed ratio of $v_\mathrm{imp}/v_\mathrm{g} \approx 0.1$ is used based on the relative speeds between the Geminids \bs grains and PSP. Assumping PSP's effective area \citep{page:20a} during this time is 6 m$^2$ and a $10-20^\circ$ cone half-angle for $\Omega$, we estimate PSP should register count rates from the Geminids in the ranges of $0.1 - 30$ hr$^{-1}$.

\section{Geminids \bs ejection geometry} \label{app:geo}

\begin{figure}
\plotone{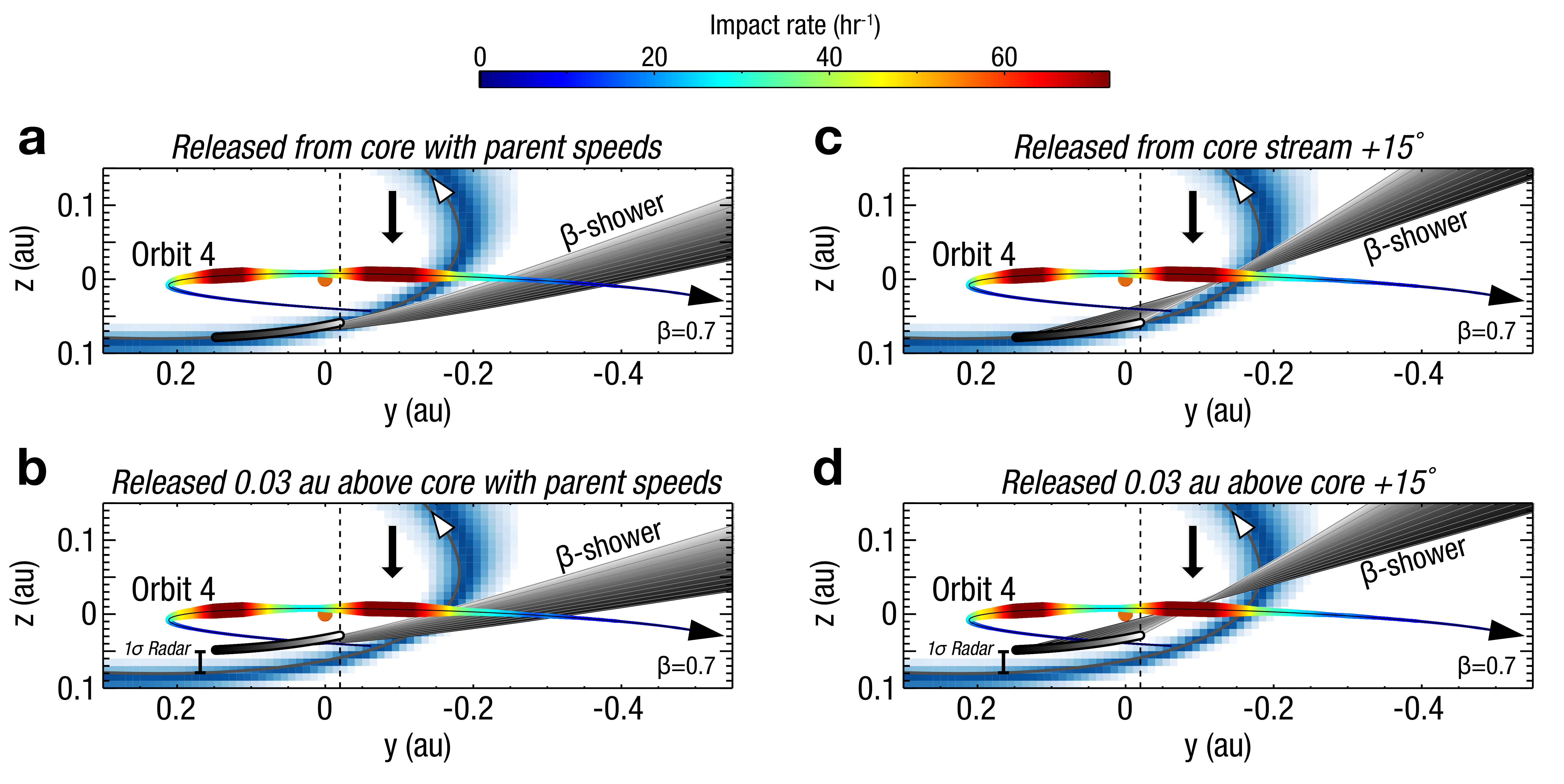}
\caption{Side view of Geminids \bs geometry. All panels show the same quantities as Figure~\ref{fig:gem} in the ecliptic J2000 y-z plane for $\beta=0.7$, with impact rates for solely orbit 4. a) Nominal case of \bmm trajectories released from the core of the Geminids stream and inheriting the initial speeds from the parent meteoroids.  b) The same trajectories as panel (a), offset by $z = +0.03$ au, $1\sigma$ of the width at this location predicted by radar meteor observations \citep{hajdukova:17a}. c) \bmm trajectories tilted by 15$^\circ$ in the $+z$ direction. d) \bmm trajectories combining both panels (b) and (c), offset by $z = +0.03$ au and with a 15$^\circ$ $+z$ cone angle.  \label{fig:gem_side}}
\end{figure}

While the trajectories of the Geminids \bs intersect the PSP orbit exactly where the post-perihelion enhancement is observed in the ecliptic x-y plane, these trajectories transit under the PSP orbit (Figure~\ref{fig:gem_side}). We investigate two effects that could enable PSP to experience fluxes from the \bsn: angular spread and a larger collisional source region.

The blue contours in Figure~\ref{fig:gem_side} show the spatial extent of the tube by integrating $\sim$14,000 orbits of visual Geminids meteors observed at Earth \citep{jenniskens:18a}. The orbits we use to map the tube structure form the more concentrated ``core'' of the stream, as they are from visual observations of larger meteoroids that would maintain orbits very similar to Phaethon. This core is also likely responsible for the narrow Geminids feature observed by PSP's WISPR imager \citep{battams:20a}. Radar observations of smaller meteoroids \citep{hajdukova:17a} more subject to additional non-gravitational perturbations indicate that 1$\sigma$ of the reconstructed width at the \bs source region discussed here is $\sim$0.03 au.

Additionally, Earth might not be representatively sampling the entirety of the Geminids tube, potentially missing a significant fraction of meteoroids that comprise a larger structure than the one we recreate here. A survey of the Geminids shower activity over the last few decades in combination with dynamical modeling shows Earth might be transiting near the edge of the stream structure \citep{ryabova:18a}. For these reasons, we expect the spatial extent to be more dispersed throughout the stream than portrayed here. We therefore offset the initial coordinates of the \bs by 0.03 au and find the shower does intersect a portion of the enhanced impact rate region along PSP's orbit (Figure~\ref{fig:gem_side}b).

Alternatively, assuming the collisional process produces fragments within a cone, we rotate the velocity vectors vertically about a radial direction by 15$^\circ$ (Figure~\ref{fig:gem_side}c) and find this also provides an intersection between the \bs and PSP enhancement region. Since we find a combination of both corrections (Figure~\ref{fig:gem_side}d) to best match the location of the enhancement, if a Geminids \bs is responsible for the post-perihelion enhancements in orbits 4 and 5, the tube could be larger than modeled here and collisional fragments could spray into a cone of $\sim$15$^\circ$.

\section{Additional \bss which intersect PSP's orbit}
In addition to investigating the Geminids, which we favor as a likely \bs candidate, we performed a comprehensive search to test if any other \bss could intersect PSP's orbit. To do so, using a model we release \bms throughout the orbits of known cometary and asteroidal bodies over a range of $\beta$ values. We then determine how near each of these \bmm trajectories transits to PSP's orbit, for this case specifically we focus on orbit 4. Table~\ref{table:bs} shows the results of this investigation, which lists the $\beta$ value and ejection location of \bms that transit nearest to PSP for the top 30 best candidates. The weight listed is the sum of squares of the difference between \bmm and PSP orbits near their closest points. Note, many of these objects are asteroidal in nature, and with the exception of Phaethon, not expected to have substantial debris trails.

\begin{sidewaystable}
\tiny
\begin{centering}
\begin{tabular}{cccccccccc}[ht!]
    Parent Body Description  &  MPC ID &  Weight &  a (au) & e & i (deg) & Lon. Asc. Node (deg) & Arg. of periapsis (deg) & Ejection true anomaly (deg) &       $\beta$ \\
\hline \hline

 2017 TC1             &   K17T01C &  0.08765 &         2.49370 &      0.96948 &           9.28577 &                 274.33215 &                   253.11950 &                  -139.04569 &    0.34 \\  
 2008 MG1             &   K08M01G &  0.18731 &         0.78321 &      0.82277 &           5.71888 &                 352.31482 &                   108.99051 &                    78.03032 &    0.96 \\  
 2019 MQ2             &   K19M02Q &  0.22405 &         1.24427 &      0.79155 &           0.86036 &                 140.51356 &                     2.58481 &                    91.72988 &    0.62 \\  
 P/2010 H3 (SOHO)     &  PK10H030 &  0.22689 &         3.06328 &      0.98549 &          23.19750 &                  77.30420 &                    25.94300 &                   128.15754 &    0.95 \\  
 2017 MM7             &   K17M07M &  0.30649 &         2.06194 &      0.96157 &          23.45918 &                 250.78365 &                   230.54887 &                   111.11971 &    0.81 \\  
 (3200) Phaethon      &     03200 &  0.32130 &         1.27137 &      0.88983 &          22.25951 &                 265.21769 &                   322.18660 &                   -45.19997 &    0.98 \\  
 2001 VB              &   K01V00B &  0.32164 &         2.34292 &      0.89387 &           8.55212 &                 299.62327 &                   230.74255 &                    73.21723 &    0.74 \\  
 322P/SOHO            &     0322P &  0.35598 &         2.50843 &      0.97976 &          11.45970 &                 351.46930 &                    56.93660 &                   151.86599 &    0.99 \\  
 2014 OX299           &   K14OT9X &  0.38085 &         2.20797 &      0.85483 &           6.45584 &                 154.49171 &                   283.03433 &                    39.52130 &    0.99 \\  
 342P/SOHO            &     0342P &  0.38731 &         3.04374 &      0.98338 &          11.97350 &                  33.80130 &                    67.41560 &                   136.49451 &    0.99 \\  
 2008 EA8             &   K08E08A &  0.38746 &         0.95730 &      0.67088 &           1.78545 &                 225.49417 &                   145.72531 &                    25.51899 &    0.29 \\  
 2011 XA3             &   K11X03A &  0.41532 &         1.46668 &      0.92597 &          28.02557 &                 273.44955 &                   323.92086 &                   -72.15717 &    0.98 \\  
 2001 QJ96            &   K01Q96J &  0.41914 &         1.59194 &      0.79758 &           5.86449 &                 338.63127 &                   121.77849 &                    33.34591 &    0.54 \\  
 2019 BE5             &   K19B05E &  0.43269 &         0.61011 &      0.65913 &           1.43626 &                 309.02706 &                     9.75573 &                   111.86369 &    0.52 \\  
 (114158) 2002 VE70   &     b4158 &  0.43386 &         1.26635 &      0.92662 &          23.77742 &                  39.58359 &                   149.56361 &                  -160.37104 &    0.36 \\  
 2015 DC200           &   K15DK0C &  0.43583 &         1.77711 &      0.86590 &           3.05844 &                 224.17269 &                    85.72899 &                  -159.15922 &    0.31 \\  
 2012 US68            &   K12U68S &  0.45501 &         2.50317 &      0.95790 &          25.80834 &                  40.49586 &                   190.74041 &                  -174.71960 &    0.29 \\  
 2019 UJ12            &   K19U12J &  0.46491 &         2.42465 &      0.94348 &          27.48587 &                 211.04715 &                    40.56684 &                   -61.02572 &    0.99 \\  
 1996 BT              &   J96B00T &  0.46584 &         1.20677 &      0.83362 &          12.10530 &                 296.85776 &                   328.23513 &                   -84.65248 &    0.95 \\  
 323P/SOHO            &     0323P &  0.50569 &         2.58246 &      0.98484 &           5.33740 &                 324.37460 &                   353.04870 &                   168.73639 &    0.95 \\  
 2015 CG13            &   K15C13G &  0.50891 &         2.50631 &      0.91341 &           6.27430 &                 124.67063 &                   238.05924 &                  -124.39949 &    0.32 \\  
 2019 YV2             &   K19Y02V &  0.53408 &         1.22723 &      0.89717 &           6.45997 &                 129.14947 &                   107.71557 &                  -110.13753 &    0.88 \\  
 2015 HG9             &   K15H09G &  0.53418 &         2.13351 &      0.84329 &           5.23836 &                 225.43773 &                   218.68841 &                    38.03775 &    0.38 \\  
 2015 KO120           &   K15KC0O &  0.55629 &         1.88183 &      0.93466 &           2.19757 &                 240.67466 &                   199.03772 &                    87.47979 &    0.98 \\  
 (195426) 2002 GU51   &     j5426 &  0.61320 &         2.53303 &      0.92217 &           6.83110 &                 233.34100 &                   200.73095 &                    53.08246 &    0.74 \\  
 2016 EL56            &   K16E56L &  0.62822 &         2.02813 &      0.85579 &          11.35514 &                 334.61334 &                    70.21392 &                    72.42659 &    0.32 \\  
 2014 JS54            &   K14J54S &  0.63082 &         2.07434 &      0.89409 &           3.36280 &                 235.05906 &                   206.53006 &                    46.05289 &    0.94 \\  
 2007 PR10            &   K07P10R &  0.64163 &         1.23182 &      0.89242 &          20.92875 &                 335.15196 &                   190.69424 &                    61.85728 &    0.40 \\  
 2013 AT72            &   K13A72T &  0.66963 &         0.67829 &      0.53565 &          15.01660 &                 118.18749 &                   167.76605 &                   -45.29174 &    0.86 \\  
 2P/Encke             &     0002P &  0.68005 &         2.21528 &      0.84800 &          11.76460 &                 334.55170 &                   186.56250 &                   -38.62496 &    0.41 \\
\hline \hline
\end{tabular}
\caption{Top 30 possible \bss for Orbit 4 \& 5 post-perihelion enhancement.\label{table:bs}}
\end{centering}
\end{sidewaystable}


\clearpage

\bibliographystyle{aasjournal}



\end{document}